\documentclass{emulateapj}

\usepackage{graphicx}
\usepackage[space]{grffile}
\usepackage{latexsym}
\usepackage{amsfonts,amsmath,amssymb}
\usepackage{url}
\usepackage[utf8]{inputenc}
\usepackage{fancyref}
\usepackage{hyperref}
\hypersetup{colorlinks=false,pdfborder={0 0 0},}
\usepackage[super]{nth}
\usepackage{natbib}
\usepackage{afterpage}

\newcommand{\civ}{\ion{C}{4}}

\newcommand{\ltwofive}{$L_{\rm 2500\,\AA}$}

\newcommand{\ebv}{{\rm E(B-V)}}
\newcommand{\dal}{\Delta(\alpha_{\lambda})}
\newcommand{\dgi}{\Delta(g-i)}
\newcommand{\dug}{\Delta(u-g)}
\newcommand{\diz}{\Delta(i-z)}
\newcommand{\mrel}{\Delta m}
\newcommand{\mbh}{M_{\rm BH}}

\shortauthors{Coleman Krawczyk}

\begin{document}
\title{Mining for Dust in Type 1 Quasars}
\shorttitle{Mining for Dust in Type 1 Quasars}  
  
\author{
Coleman M.\ Krawczyk,\altaffilmark{1,2}%\altaffilmark{2}
Gordon T.\ Richards,\altaffilmark{1,3}
S. C. Gallagher,\altaffilmark{4,5}
Karen M. Leighly,\altaffilmark{6} 
Paul C.\ Hewett,\altaffilmark{7}
Nicholas P.\ Ross,\altaffilmark{1}
P. B. Hall,\altaffilmark{8}
}
\altaffiltext{1}{Department of Physics, Drexel University, 3141 Chestnut Street, Philadelphia, PA 19104, USA}
\altaffiltext{2}{Institute for Cosmology and Gravitation, University of Portsmouth, Dennis Sciama Building, Burnaby Road, Portsmouth, PO1 3FX, UK}
\altaffiltext{3}{Max-Planck-Institut f{\"u}r Astronomie, K{\"o}nigstuhl 17, D-69117 Heidelberg, Germany}
\altaffiltext{4}{Department of Physics \& Astronomy, The University of Western Ontario, London, ON N6A 3K7, Canada}
\altaffiltext{5}{Visiting Fellow, Yale Center for Astronomy and Astrophysics, Yale University, P.O. Box 208120, New Haven, CT 06520, USA}
\altaffiltext{6}{Homer L. Dodge Department of Physics and Astronomy, The University of Oklahoma, 440 W. Brooks Street, Norman, OK 73019, USA}
\altaffiltext{7}{Institute of Astronomy, University of Cambridge, Madingley Road, Cambridge CB3 0HA, UK}
\altaffiltext{8}{Department of Physics and Astronomy, York University, Toronto, ON M3J 1P3, Canada}

\begin{abstract}
We explore the extinction/reddening of $\sim$35,000 uniformly selected quasars with $0<z\leq5.3$ in order 
to better understand their {\em intrinsic} optical/ultraviolet (UV) spectral energy distributions.
%With a sample of $\sim35,000$ uniformly selected quasars with redshifts $0<z\leq5.3$, 
%we explore their extinction/reddening in order to better understand their {\em intrinsic} SEDs. 
Using rest-frame optical--UV photometry taken from the Sloan Digital Sky Survey's (SDSS) 7th data release, cross-matched to
{\em WISE} in the mid-infrared, 2MASS and UKIDSS in the near-infrared, and {\em GALEX} in the UV, 
we isolate outliers in the color distribution and find them well described by an SMC-like reddening law.
A hierarchical Bayesian model with a Markov Chain Monte Carlo sampling method was used to find distributions of powerlaw indices and $\ebv$ consistent with both 
the broad absorption line (BAL) and non-BAL samples. We find that, of the $ugriz$ color-selected type 1 quasars in SDSS, 
2.5\% (13\%) of the non-BAL (BAL) sample are consistent with $\ebv>0.1$ and 0.1\% (1.3\%) with $\ebv>0.2$. Simulations show both populations of quasars are intrinsically bluer than the mean composite, with a mean spectral index ($\alpha_{\lambda}$) of $-1.79$ ($-1.83$).  
The emission and absorption-line properties of both samples reveal that quasars with intrinsically red continua have narrower Balmer lines and stronger high-ionization emission lines, the latter indicating a harder continuum in the extreme-UV and the former pointing to differences in black hole mass and/or orientation.
% or a multi-component broad emission line region.
%indicating either smaller BH mass or more face-on orientation.
\end{abstract}


\section{Introduction} \label{sec:dust_intro}

For distant quasars, several mechanisms contribute to extinction: the Milky Way (MW), galaxies along the line of sight, the quasar host galaxy, and the nuclear region itself.  While it is relatively straightforward to correct for and/or mitigate against the first two sources of reddening, it can be quite difficult to disentangle the last two sources and to properly isolate the effects of reddening from intrinsic differences in the spectral energy distribution (SED). Indeed, quasars can appear ``red'' (or rather, redder than average) both because they are intrinsically red or because of dust reddening and there is a growing body of work that broadly addresses this issue \citep[e.g.,][]{Webster:1995,Francis:2000,Richards:2003,Hopkins:2004,Davis:2007,Banerji:2012,Glikman:2012,Krogager:2015}.

\iffalse
Quasar SEDs between the ultraviolet (UV) and near-infrared (near-IR) are generally described by a modified black body \citep{Shakura:1973} which is produced by an accretion disk with a range of temperatures.  At wavelengths which start to sample the high-temperature limit of the accretion disk, the SED begins to ``turn over,'' following the Wien part of the distribution.  The location of this turnover is likely a function of luminosity \citep[e.g.,][]{Scott:2004,Shang:2005,Davis:2007}. At longer wavelengths, contamination from the host galaxy itself can make the intrinsic contribution of the central engine difficult to measure. Once the host galaxy contribution is accounted for, it is generally found that, between the rest-frame wavelengths of 1\,$\mu$m and 1216\,\AA, a quasar's continuum can be roughly approximated as a powerlaw \citep[e.g., ][]{Vanden-Berk:2001,Richards:2006,Krawczyk:2013}. 
\fi

The ultraviolet (UV) through near-infrared (near-IR) SEDs of quasars are thought to be produced by radiation from a geometrically thin, optically thick accretion disk \citep{Shakura:1973} (SS73).  Once the host galaxy contribution is accounted for, it is generally found that, between the rest-frame wavelengths of 1\,$\mu$m and 1216\,\AA, a quasar's continuum can be roughly approximated as a powerlaw \citep[e.g., ][]{Vanden-Berk:2001,Richards:2006,Krawczyk:2013}.  In reality some curvature of the SED is expected in even the simplest multi-temperature black body approximation due to a longer wavelength turn over of the highest-temperature black body peak of the accretion disk SED for (all else being held equal): higher mass, lower accretion rate, lower spin, and face-on orientations\citep{Hubeny:2000,Davis:2011}.  This change in spectral index (curvature of the SED) is seen empirically and has been found to be a function of luminosity \citep[e.g.,][]{Scott:2004,Shang:2005,Davis:2007}.  More complex treatment of accretion disk SEDs \citep[e.g.][]{Hubeny:2000} also leads to deviations from power-law continua even well away from the UV peak of the SED.

\citet{Koratkar:1999}, \citet{Davis:2011}, and \citet{Capellupo:2015} and references therein provide a summary of attempts to reconcile the SS73 thin accretion disk model with observations of quasars.  We do not consider deviations from the thin accretion disk model herein, but such deviations could be important, particularly for luminous quasars that have high (mass-weighted) accretion rates, which could lead to the formation of a slim, rather than thin accretion disks \citep{Abramowicz:1988,Netzer:2014}.

Investigating the detailed dependence of the SED on black hole physics
is beyond the scope of this paper; however, we can investigate the
effect of dust on the quasar SED by assuming that the Wien part of the
black body is predominantly at higher frequency than our observations,
and describing the continuum by:
\begin{equation} \label{eqn:red_eq} 
L_{\lambda,{\rm rest}} \propto \lambda^{\alpha_\lambda} e^{-\tau_\lambda} \propto \lambda^{\alpha_\lambda} 10^{-{\rm E(B-V)}R_{\lambda}/2.5} 
\end{equation}
where $L_{\lambda,{\rm rest}}$ is the rest frame luminosity, $\alpha_\lambda$ is the intrinsic spectral index, $\tau_\lambda$ is the optical depth of the dust, and $R_{\lambda}$ is a function that is dependent on the physical properties of the dust.  For small amounts of dust reddening, the SED will simply appear to have a ``steeper'' (redder) powerlaw and will be observationally degenerate with changes to the thin accretion disk parameters.  For larger amounts of reddening, noticeable curvature can be seen and more clearly distinguished from changes in the intrisic SED of the disk.  In Section~\ref{sec:red_v_red:phot} we use this curvature as in indicator for the type of dust present in quasars.

\iffalse
However, if there is a significant amount of dust associated with the
quasar, then the continuum will take on the form:
\begin{equation} \label{eqn:red_eq} 
L_{\lambda,{\rm rest}} \propto \lambda^{\alpha_\lambda} e^{-\tau_\lambda} \propto \lambda^{\alpha_\lambda} 10^{-{\rm E(B-V)}R_{\lambda}/2.5} 
\end{equation}
where $L_{\lambda,{\rm rest}}$ is the rest frame luminosity, $\alpha_\lambda$ is the intrinsic spectral index, $\tau_\lambda$ is the optical depth of the dust, and $R_{\lambda}$ is a function that is dependent on the physical properties of the dust.  For small amounts of dust reddening, the SED will simply appear to have a ``steeper'' (redder) powerlaw.  For larger amounts of reddening, noticeable curvature can be seen.  In Section~\ref{sec:red_v_red:phot} we use this curvature as in indicator for the type of dust present in quasars.
\fi

Deviations from the mean quasar SED have been used previously to define red quasar samples, including \citet{Gregg:2002} who used an optical-infrared spectral index of $\alpha<-1$ as their defining property. Very dust reddened quasars can be difficult to identify in optical surveys, both because of their redder than average color (often similar to stellar colors) and the corresponding extinction. Radio and infrared selection \citep[e.g., ][]{Gregg:2002,Glikman:2007,Maddox:2008,Maddox:2012,Banerji:2012,Fynbo:2013,Ross:2014} help to overcome this. These studies suggest that a significant portion ($>25\%$ --- depending on the flux limit of and wavelength selection) of the underlying quasar population is dust reddened (yet still classified as Type 1 by virtue of having broad emission lines).

The ability to identify the nature of any dust reddening, determine the magnitude of the effect, and correct for it is important for a number of reasons, for example, determining the true spectral index and optical luminosity of the quasars, determining how much of the IR continuum could be emitted from an optically thin (as opposed to optically thick dust) dust component, and, crucially, determining the {\em bolometric} luminosity of the quasar \citep{Runnoe:2012}.  The bolometric correction is crucial for making comparisons between quasars photometered at different rest-frame wavelengths, the extreme example being comparisons of unobscured (type 1) and obscured (type 2) quasars \citep[e.g., ][]{Zakamska:2003,Merloni:2014}, where the luminosity of the latter must be measured not in the (obscured) optical/UV continuum, but instead from narrow emission lines, the IR or X-ray. Accurate bolometric luminosities are also needed for accurate estimates of the accretion rate and the Eddington ratio, $L_{\rm Bol}/L_{\rm Edd}$. Thus proper determination of the SED provides a powerful link from observed properties to the parameters of accretion disk physics: mass, accretion rate, and spin \citep[SS73;][]{Hubeny:1997}.

In this paper we aim to identify the mean type of dust extinction present in the observed population of quasars and to construct de-reddened quasar SEDs for a subset of the Sloan Digital Sky Survey's (SDSS) \nth{7} data release (DR7) quasar catalog \citep{Schneider:2010}.  By distinguishing between objects that are intrinsically red and those that are dust reddened, we can also comment on the broad emission \citep[e.g., ][]{Richards:2003} and broad absorption line (BAL) \citep[e.g., ][]{Reichard:2003a,Reichard:2003b,Baskin:2013} properties of quasars as a function of their intrinsic color (and the amount of dust reddening).  These emission line properties can help to reveal the SED shape outside of the range covered by our data, as a function of the masses, and/or inclinations that correlate with intrinsic color.
To this end this paper is structured as follows:  the data is presented in Section~\ref{sec:data}.  Section~\ref{sec:model} focuses on the identification and fitting of various dust reddening models including individual reddening values for each quasar in our sample. We analyze composite spectra as a function of these fitted values in Section~\ref{sec:spectra} and present our conclusions in Section~\ref{sec:conclusions}.

\section{Data} \label{sec:data}

For our data set we use the mid-infrared (mid-IR) through UV cross-matched catalog provided by \citet{Krawczyk:2013}.  These data are taken from the SDSS DR7
%Sloan Digital Sky Survey's (SDSS) 7th data release (DR7) 
quasar catalog \citep{Schneider:2010} and have been cross-matched to {\em Spitzer} and {\em WISE} \citep[Wide-field Infrared Survey Explorer,][]{WISE} in the mid-IR, 2MASS \citep{2MASS} and UKIDSS \citep[UKIRT Infrared Deep Sky Survey,][]{UKIDSS} in the near-IR, and {\em GALEX} \citep[Galaxy Evolution Explorer,][]{GALEX} in the UV. The host galaxy component for each quasar was estimated using the \citet{Fioc:1997} elliptical galaxy template with a normalization estimated by combining two different models: the relation outlined in \citet{Shen:2011} for the high luminosity quasars, and the relationship from \citet{Richards:2006} for the low luminosity quasars \citep[see Section~3.3 of][]{Krawczyk:2013}. To mitigate against any biases due to selection effects, we limit our study to the uniformly selected point source quasars \citep{Richards:2006a}, leaving an initial sample of 55,772 quasars.  
We remove the effects of dust reddening/extinction within the MW itself by applying the corrections of \citet{Schlegel:1998} and \citet{Schlafly:2011}.  

%For this analysis we limit our sample to the uniformly selected point source quasars \citep{Richards:2006a}, leaving an initial sample of 55,772 quasars. This cut mitigates against any biases due to selection effects.

In order to better understand the dust present at a quasar's redshift, we further desire to minimize the effects of reddening caused by intervening galaxies.  We achieve this by removing 16,450 quasars that show signs of strong absorption line systems along the line of sight.
The presence of strong absorption lines in the spectrum of a quasar generally indicates that the light is passing through an intervening galaxy.  As dust in those galaxies can redden the quasar spectrum \citep{York:2006,Menard:2008,Khare:2012}, it is important to remove such quasars from our sample.
We specifically removed all quasars with at least two identified absorption lines (grade C or higher in the parlance of \citet{York:2005}) with a \ion{Mg}{2} line having an equivalent width (EW) $\geq0.3$\,\AA\ (York et al. 2015, in preparation).   These systems are thought to be associated with absorbing columns of $N_{\rm HI} \sim 10^{18-20} {\rm cm^{-2}}$ \citep{Churchill:2005,Prochaska:2010}.

This process yields a sample of quasars that is not selected preferentially because of radio properties and whose observed SED should be dominated by the physics of the accretion disk and dust reddening/extinction that is local to the quasar and/or the host galaxy.  As BAL quasars are seen to have different reddening properties \citep{Sprayberry:1992} and perhaps intrinsically different SEDs \citep{Weymann:1991,Reichard:2003b}, we split the sample into two groups: BAL quasars, as
identified by \citet{Shen:2011} and York et al. (2015, in preparation), and non-BAL quasars, containing 1744 and 34,233 quasars respectively.  Our goal is to investigate the dust properties of otherwise normal type 1 quasars as opposed to specifically investigating dusty quasars \citep[e.g.,][]{Fynbo:2013,Ross:2014}.

\begin{figure}[t]
\begin{center}
\includegraphics[width=3in]{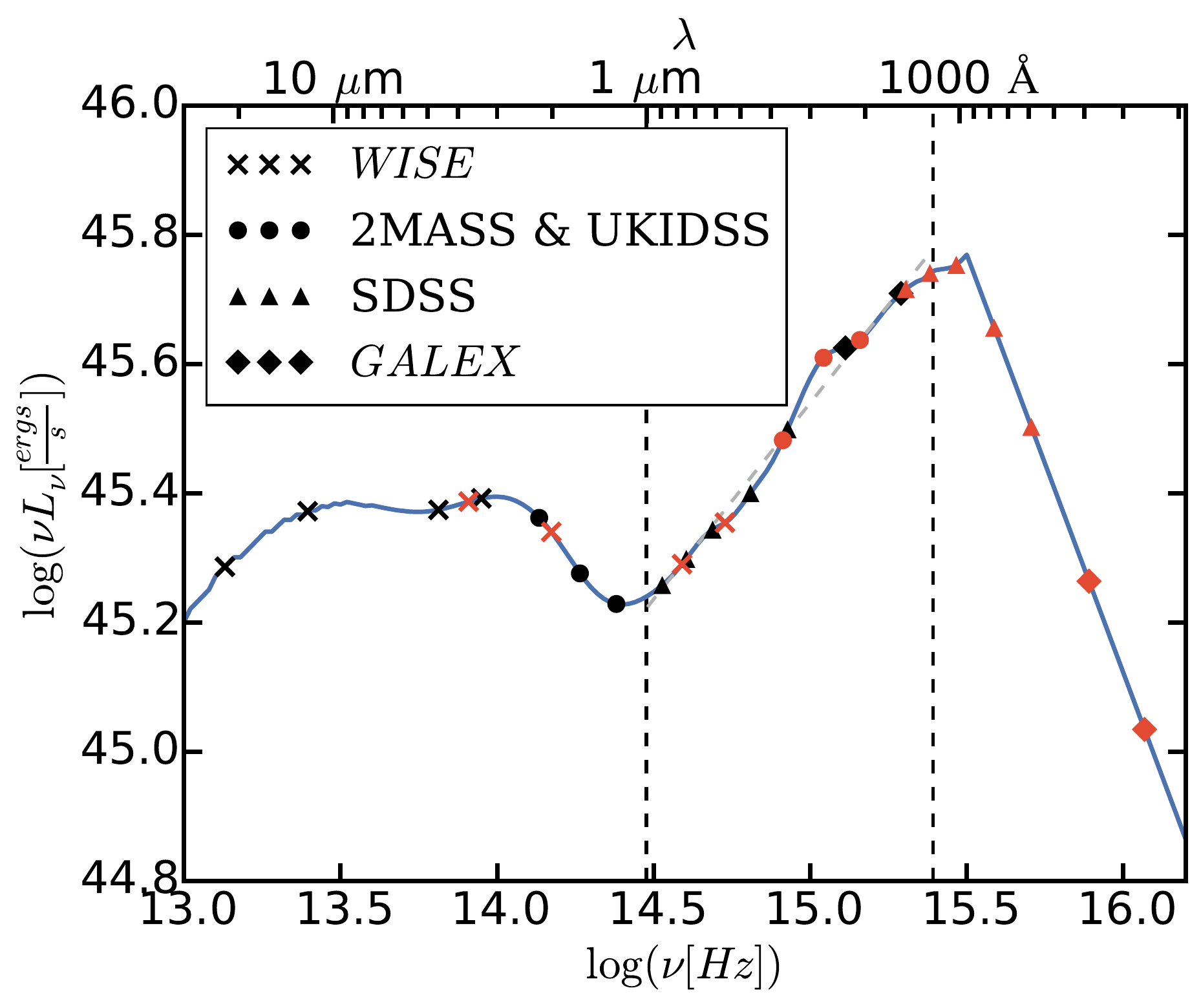}
\caption[Filter position vs. redshift]{\label{fig:sed_redshift} Position of various filters on the mean quasar SED (with the major spectral lines removed) for $z=0$ (black) and $z=5$ (orange).  The filters are from {\em WISE} (x's), 2MASS/UKIDSS (circles), SDSS (triangles), and {\em GALEX} (diamonds).  The vertical dashed lines show the range where the SED is well approximated as a powerlaw (dashed gray line $\alpha_\nu=-0.39$); all but the two reddest filters pass through this range for the redshift distribution of our sample.}
\end{center}
\end{figure}

For our dust analysis we only use the filters that fall between 1\,$\mu$m and 1216\,\AA\ in a quasar's rest frame (the portion consistent with a powerlaw). Because of these limits, only the following filters are used: {\em NUV} and {\em FUV} from {\em GALEX}, {\em ugriz} from SDSS, {\em JHK} from either 2MASS or UKIDSS, and {\em W1-2} from {\em WISE}.  Figure~\ref{fig:sed_redshift} shows the positions of these filters on the mean quasar SED from \citet{Krawczyk:2013} for $z=0$ (black) and $z=5$ (orange).  This mean SED has a powerlaw index of $\alpha_\nu=-0.39$ (gray dashed line) and the small blue bump (SBB) is visible at $\sim3500$\,\AA.

\section{Fitting Reddening Models to Quasar Photometry} \label{sec:model}

\begin{figure*}%[ht]
\begin{center}
\includegraphics[width=6in]{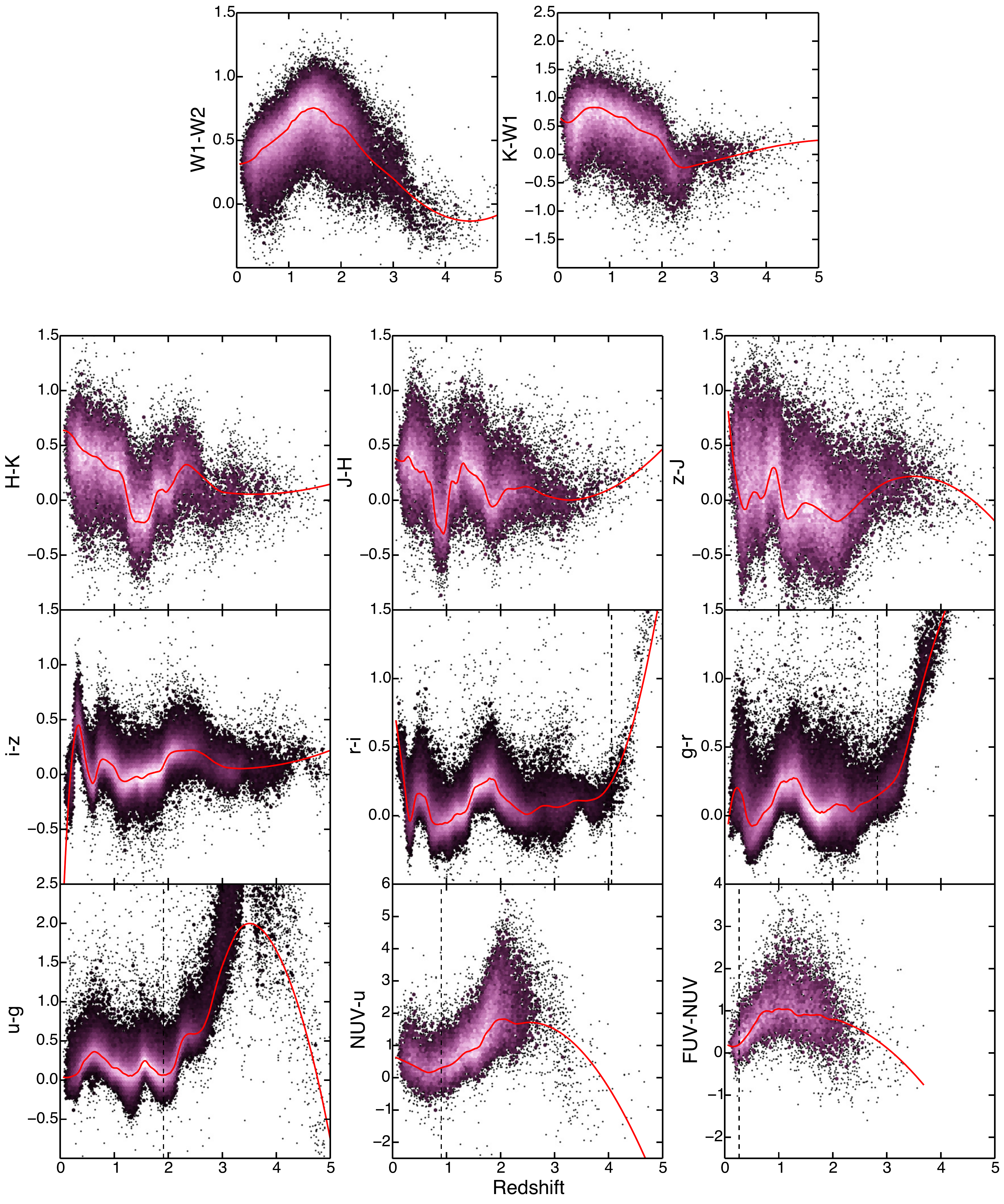}
\caption[Colors vs. redshift]{\label{fig:mode} Colors as a function of redshift with the modes indicated by the solid red lines.  The colors indicate the linear density of points with white being most dense and the outliers shown as black points. The vertical dashed lines indicate where the bluest filter for each color passes into the Ly$\alpha$ line.}
\end{center}
\end{figure*}

\subsection{Modal Colors}

To identify a quasar as ``red,'' we must first determine both
the expected distribution of SEDs and a method for identifying which
quasars can be considered as outliers from that distribution.
The expected observed SED changes as a function of redshift as the different spectral lines are shifted though the filter sets.  By examining quasar colors as a function of redshift we can find the typical quasar SED, and assuming this SED contains little or no dust, we can use it to characterize the effects that different reddening laws have on the SED. We note that these colors are calculated after removing the estimated host galaxy contribution (most relevant at z$\lesssim$0.8) using the prescription discussed in \citet{Krawczyk:2013}.

Previous studies have mitigated against emission lines changing a quasar's colors by binning the data by redshift and removing the median \citep[e.g.,][]{Richards:2003} or mode \citep[e.g.,][]{Hopkins:2004} within each bin. The results of this process are strongly dependent on the bin width chosen.  To avoid this issue, we used a non-parametric \nth{2} degree local regression to estimate the modal color as a function of redshift.  We used the \texttt{loess} regression package in \texttt{R} \citep{R:2014} with smoothing parameters chosen so the regression followed the ridge-line (i.e. the mode) for each color-redshift plot, as in Figure~\ref{fig:mode}.  Although this method does not work well once a filter passes into the Lyman forest 
%(e.g. $u-g$ for $z\gtrsim2$)
(vertical dashed lines in Figure~\ref{fig:mode}), with the rest-frame wavelength cuts applied in the analysis, this discrepancy does not affect our results.

\begin{figure*}%[ht]
\begin{center}
\includegraphics[width=6in]{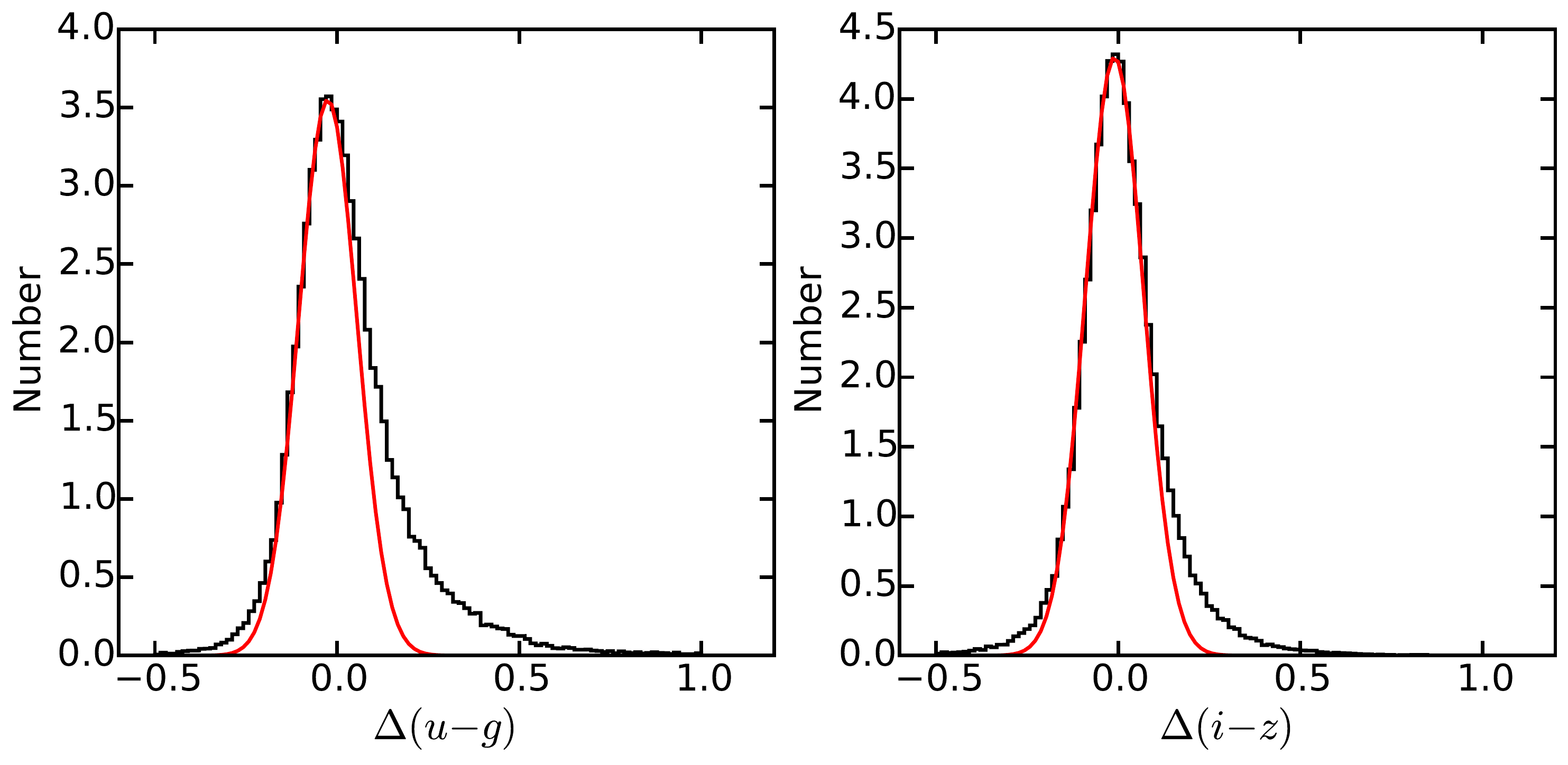}
\caption[Relative colors]{\label{fig:Dug} 
%Relative colors. 
Relative colors for $\dug$ ({\em left}) and $\diz$({\em right}). $\dug$ shows a heavy red tail as compared to its best fit Gaussian (red), while $\diz$ shows a weaker red tail as compared to its best fit Gaussian. This asymmetry indicates that there is a population of dust reddened quasars in our sample.}
\end{center}
\end{figure*}

With these modal colors, we define relative colors \citep{Richards:2003} as the difference between the observed color and the modal color at the quasar's redshift (e.g. $\dug=(u-g)-\langle u-g \rangle_{{\rm mode},z}$).  From these relative colors we can also build relative magnitudes ($\mrel$) up to a normalization factor; we adopt $\mrel_{i{\rm band}}=0$ throughout this paper.  These relative colors and magnitudes allow for a color analysis that is independent of emission line effects.
%for which it is possible to make fair comparisons of different colors.

While we discuss the optical/UV continuum in terms of reddening an intrinsic power-law, this process preserves the modal continuum of quasars regardless of its shape.  That is, it does not matter if the true SED has some intrinsic curvature as discussed in Section~1.  The process does however suffer from the usual problem of short-wavelengths being dominated by high-$z$ quasars and long wavelengths by low-$z$ quasars.

To determine if we have a population of quasars affected by dust reddening, we compare the distributions for a relative color on the blue end of the quasar SED ($\dug$) and and relative color on the red end of the quasar SED ($\diz$).  If there is no dust reddening in our population, we would expect the shape of these two distributions to be the same, whereas dust reddening would cause there to be (relatively) more red quasars on the blue end of the SED than the red end of the SED.  Thus we can distinguish between dust reddening and synchrotron emission extending to the optical since dust preferentially absorbs the shorter-wavelength photons (e.g. adding a heavy red tail to $\dug$), while synchrotron adds long-wavelength photons (e.g. adding a heavy red tail to $\diz$); see \citet{Francis:2000}.  Figure~\ref{fig:Dug} shows these two distributions and we can see that $\dug$ shows a heavier red ``tail'' than $\diz$, indicating that we do indeed have a population of dust reddened quasars.

\subsection{Red vs. Reddened: Photometry} \label{sec:red_v_red:phot}

As discussed in \citet{Richards:2003}, ``red" quasars can be intrinsically red or dust reddened.  In order to tell these two apart, we need to understand how dust changes the observed $\mrel$.  Combining Equation~\ref{eqn:red_eq} with our definition of $\mrel$ and assuming the modal quasar contains no dust\footnote{This technically measures dust reddening relative to the modal quasar and is in addition to any reddening of the modal quasar itself.}:
\begin{align} \label{eqn:rel_red_eq} 
\mrel&=m - \langle m \rangle_{{\rm mode},z} &&{\rm  (observed)} \nonumber \\
&=m_0 + 2.5\dal \log{(\lambda)} + \ebv R_\lambda &&{\rm  (model)}
\end{align}
where $m_0$ is a normalization term, $\dal$ is the difference between the quasar's {\em intrinsic} spectral index and the {\em observed} modal spectral index of $\alpha_{\lambda} =-1.72$ ($\alpha_{\nu}=-0.28$, as determined from the modal colors in Figure~\ref{fig:mode}), with positive values being bluer, and $R_{\lambda}$ is a function dependent on the properties of the dust.  Throughout the remainder of the article we will be working with values of $\dal$ since that is the term that appears in the model.  These values can be converted to absolute spectral indices ($f_{\lambda} \propto \lambda^{\alpha_{\lambda}}$, $f_{\nu} \propto \nu^{\alpha_{\nu}}$) by adding in the observed modal spectral index:
\begin{eqnarray}
	\alpha_{\lambda} &=& -1.72 - \dal \label{eqn:dal_2_al} \\
	\alpha_{\nu} &=& -2-\alpha_{\lambda} \nonumber \\
	&=& -0.28 + \dal \label{eqn:dal_2_anu}
\end{eqnarray}

\begin{figure}[t]
\begin{center}
\includegraphics[width=3in]{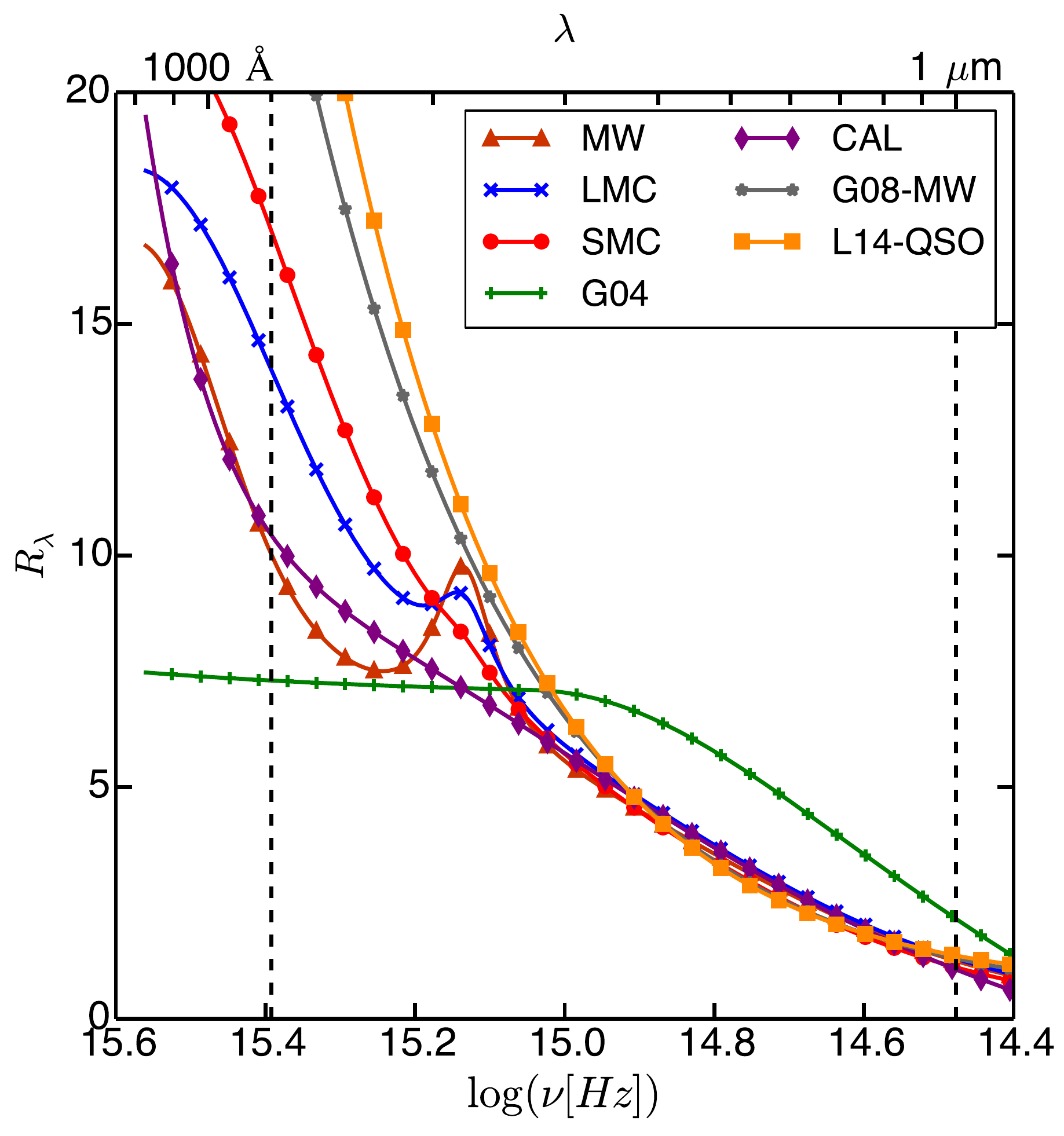}
\caption[Dust extinction curves]{ \label{fig:ecurves} Extinction curves for seven different reddening laws with $\ebv=1$: Milky Way type (brown triangles), LMC type (blue X's), SMC type (red dots), starburst type (purple diamonds), \citet{Goobar:2008} type (gray stars), \citet{Leighly:2014} type (orange squares), and \citet{Gaskell:2004} type (green +'s). The vertical dashed lines indicate the polynomial fitting range used in this work. In all cases we see dust reddening adds curvature to the relative magnitudes.}
\end{center}
\end{figure}

\begin{figure*}%[ht]
\begin{center}
\includegraphics[width=6in]{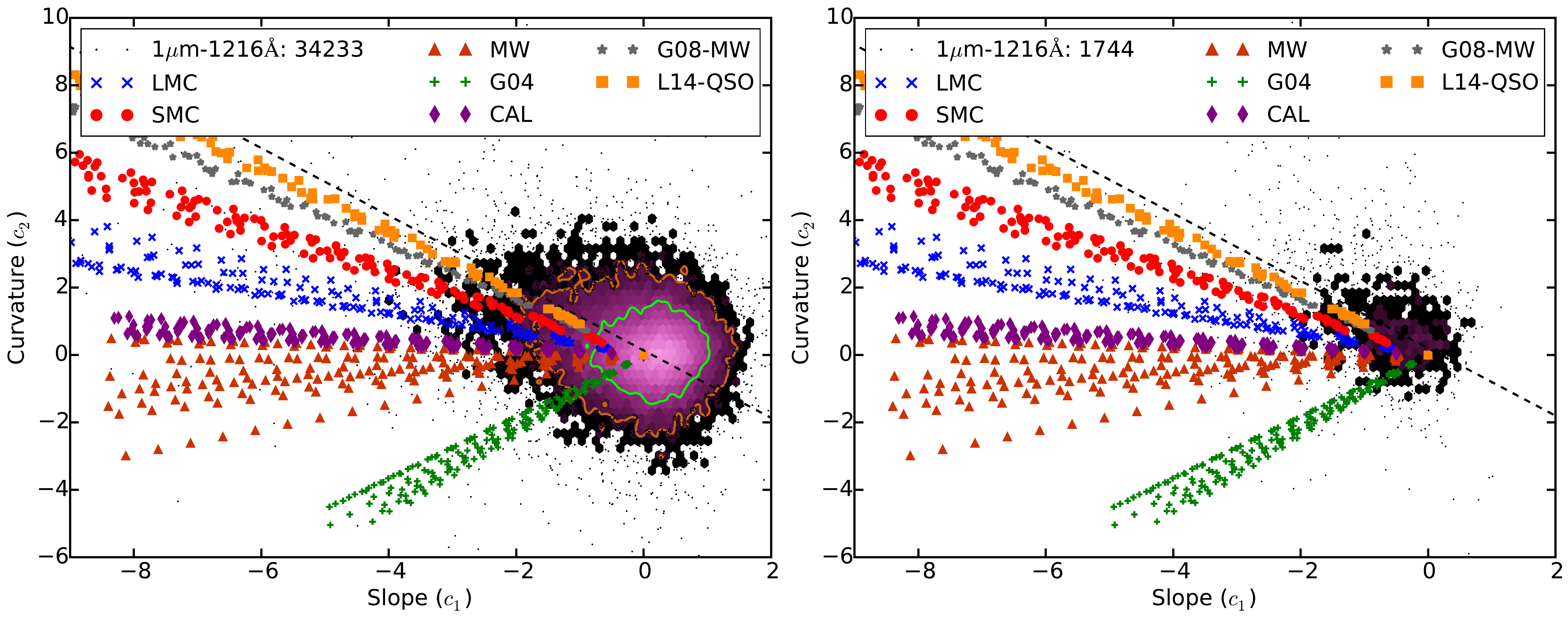}
\caption[$c_1$ vs. $c_2$]{ \label{fig:c1c2} $c_1$ vs. $c_2$ for uniformly selected SDSS quasars.  The colors indicate the linear density of scatter points with light purple being the most dense and the outliers are shown as black scatter points.  The 1 and 2$\sigma$ contours are also shown in the left panel.  The symbols show a typical quasar placed at redshifts ranging from 0 to 2.2 and with $\ebv$ values ranging from 0 to 0.8 for seven different reddening laws: SMC type (red dots), LMC type (blue X's), starburst type (purple diamonds), Milky Way type (brown triangles), multiple-scattering dust fit to the Milky Way \citep[gray stars][]{Goobar:2008}, multiple-scattering dust fit to a BAL quasar \citep[orange squares][]{Leighly:2014}, and \citet{Gaskell:2004} type (green +'s).  The black dashed line shows the orthogonal regression of the data.  {\em Left}: sample of 34,233 non-BAL quasars. {\em Right}: sample of 1,744 BAL quasars.  In both samples the trend of the data best matches the steeper reddening laws.}
\end{center}
\end{figure*}

\begin{figure*}[t]
\begin{center}
\begin{tabular}{c}
	\includegraphics[width=6in]{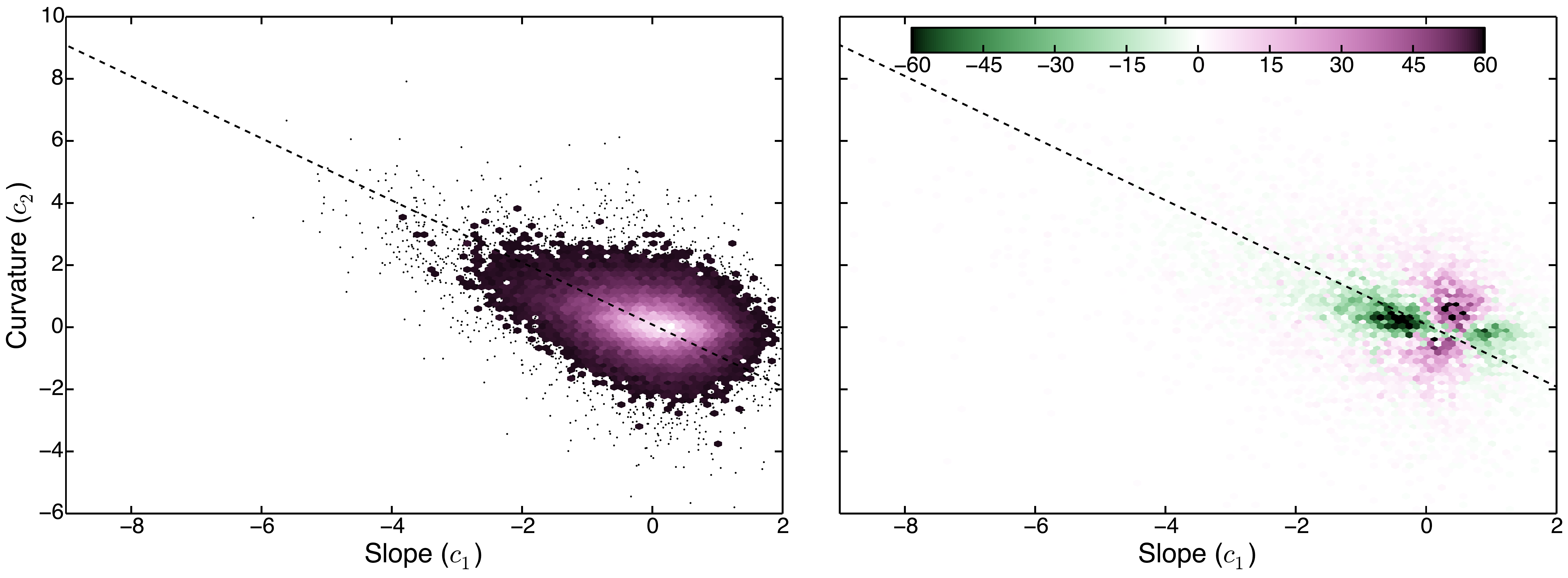} \\
	\includegraphics[width=6in]{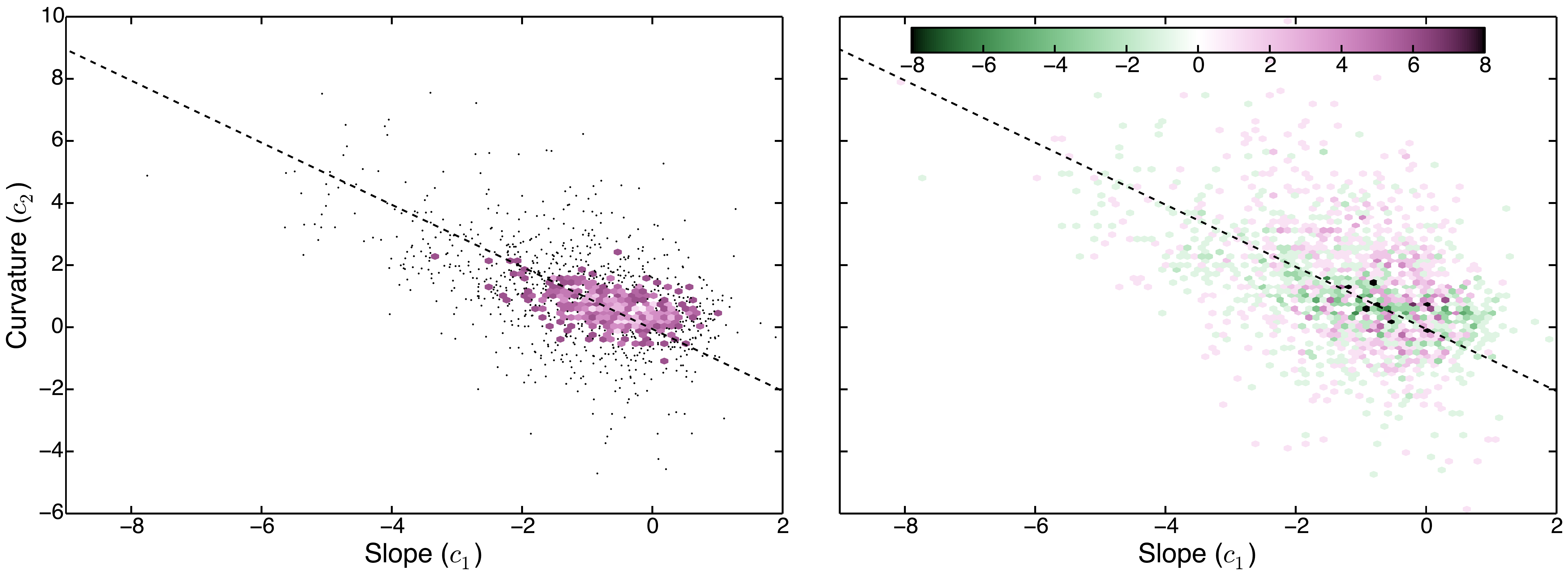}
\end{tabular}
\caption[$c_1$ vs. $c_2$ for MC sample of non-BAL quasars]{\label{fig:mcmc_no_bal_mc} 
Monte Carlo simulations of $c_1$ and $c_2$ ({\em left}) and residuals with the data ({\em right}) for the SMC reddening law based on the population parameters shown in Figures~\ref{fig:mcmc_no_bal_tri} \& \ref{fig:mcmc_bal_tri} for the non-BAL ({\em top}) and BAL ({\em bottom}) samples. The orthogonal regression for the MC distributions are shown as a black dashed line in each panel.
Our model over-predicts the number of quasars with low curvature, indicating it is likely under-predicting the amount of dust in the sample or that the powerlaw model is not a perfect representation of the continuum.}
\end{center}
\end{figure*}

As stated in Section~\ref{sec:dust_intro}, this model only holds between 1\,$\mu$m and 1216\,\AA, since that is where the underlying SED is well approximated by a single power law.  The one thing $R_{\lambda}$ has in common across all forms of dust is that it adds {\em curvature} to $\mrel$ by preferentially removing shorter wavelength photons (except for \citealt{Gaskell:2004} reddening which introduces curvature in the opposite sense).
%preferentially removes shorter wavelength photons \textbf{[GTR: except Gaskell]}, effectively adding   
Figure~\ref{fig:ecurves} shows $R_{\lambda}$ for seven different reddening laws with $\ebv=1$:  MW \citep{Pei:1992}, Large Magellanic Cloud \citep[LMC;][]{Pei:1992}, Small Magellanic Cloud \citep[SMC;][]{Pei:1992}, starburst \citep[CAL;][]{Calzetti:1995}, multiple-scattering dust models \citep[G08-MW,L14-QSO;][]{Goobar:2008,Leighly:2014}, and \citet[G04;][]{Gaskell:2004} type reddening.

From Equation~\ref{eqn:rel_red_eq} we see that changes in the intrinsic spectral index, $\dal$, will change the slope of $\mrel$ (i.e. it is only multiplied by $\log{(\lambda)}$), while changes in the amount of dust, measured by  $\ebv$, will change both the slope and curvature of $\mrel$ since $R_{\lambda}$ contains higher orders of $\log{(\lambda)}$.  \citet{Hopkins:2004} characterized these changes by fitting $\mrel$ with a \nth{1}-order Chebyshev polynomial to obtain the slope ($c_1$) and with a \nth{2}-order Chebyshev polynomial to obtain the curvature ($c_2$).  Because of the orthogonality properties of the Chebyshev polynomials, $c_1$ and $c_2$ are linearly independent of each other.  This property means that changes in a quasar's intrinsic spectral index will {\em only} change $c_1$, but given the nature of $R_{\lambda}$, changes in $\ebv$ will change {\em both} $c_1$ and $c_2$.

Figure~\ref{fig:c1c2}  shows the $c_1$ vs. $c_2$ parameter space for both the non-BAL ({\em left}) and BAL ({\em right}) samples, where the symbols show the modal quasar ($c_1$=0, $c_2$=0 by definition) reddened by seven different reddening laws with redshifts ranging from 0 to 2.2 and $\ebv$ values ranging from 0 to 0.8. If the reddening were instead applied to a quasar that was intrinsically blue (red) then the symbols would shift to higher (lower) $c_1$ values while the $c_2$ values would remain unchanged.  Looking at the trend of the BAL and non-BAL quasar samples in Figure~\ref{fig:c1c2}, it appears that the data is most closely matched with the steeper reddening laws, e.g., the SMC dust law\footnote{The SMC type dust is typically associated with having a smaller grain size than LMC or MW type dust \citep{Pei:1992}.} or the multiple-scattering dust model described by \citet{Goobar:2008} to explain anomalous dust extinction in supernovae host galaxies \citep[see also][]{Fynbo:2013,Leighly:2014}.

\subsection{Monte Carlo Parameter Estimation} \label{sec:mcmc}
One of the difficulties of our analysis is the degeneracy between changes in color and dust reddening \citep[e.g., ][]{Reichard:2003a,Reichard:2003b,Richards:2003}.  As such, to fit for both the individual $\dal$ and $\ebv$ values along with the shapes of the distributions they come from, we use a multi-level or hierarchical Bayesian model.\footnote{See \citet{Stern:2012}, Section~3.7.3 for a discussion of using the ratio between UV luminosity and the broad component of the H$\alpha$ emission line to break the degeneracy.}  In our model we assume the $\dal$ values come from a normal distribution with mean $\mu_\alpha$ and width $\sigma_\alpha$ and the $\ebv$ values come from an exponentially modified Gaussian (EMG) distribution.  The EMG is the result of summing a normal random variable with mean $\mu_{\rm dust}$ and width $\sigma_{\rm dust}$ and an exponential random variable with rate parameter $\lambda_{\rm dust}$ (smaller values indicate a heavier red tail).  The EMG was chosen since it is very similar in shape to the half-normal, half-exponential distribution \citet{Hopkins:2004} found to work well for the $\ebv$ distribution of quasars. In addition to these parameters we also assume a noise term with zero mean and width $S$ to model error in determining the modal colors.
The technical details of this process are included in Appendix~\ref{ch:mcmc} for the interested reader; we summarize the results in Section~\ref{sec:mcmc_results} and continue our analysis in Section~\ref{sec:spectra}.   See \citet{Capellupo:2015} for another example of such analysis, there attempting to model quasar SEDs as a function of black hole parameters.

\subsection{MCMC Results: Population Parameters} \label{sec:mcmc_results} 

Using the model described in Appendix~\ref{sec:mcmc_model}, we fit for both individual $\dal$ 
%\footnote{see Equations~\ref{eqn:dal_2_al} and \ref{eqn:dal_2_anu} for conversion to absolute spectral indices} 
and $\ebv$ values ($\phi_i=\{\dal_i, \ebv_i \}$ from Appendix~\ref{ch:mcmc}) and the population parameters ($\theta=\{ \mu_{\alpha}, \sigma_{\alpha}, \mu_{\rm dust}$, $\sigma_{\rm dust} ,\lambda_{\rm dust}, S \}$ from Appendix~\ref{ch:mcmc}) describing the distributions of these variables.  This method was used for both the SMC reddening law and the multiple-scattering reddening law from \citet{Goobar:2008} calibrated to fit Mrk 231 \citep[a heavily reddened BAL quasar;][]{Leighly:2014} as those two models appear to better represent the data in Figure~\ref{fig:c1c2}.

To check how well our model can recreate the observed data
%To see how well these population parameters represent our data 
we simulate an MC sample of quasars using the results of our fit to the SMC law\footnote{The results for the L14-QSO fit are qualitatively similar.} to find $c_1$ and $c_2$ values.  The resulting distributions can be seen in the left panels of Figure~\ref{fig:mcmc_no_bal_mc} while right panels show the residual with the original data. In the residual plots, green shows where the model over-estimated and purple shows where it under-estimated the number of objects.  In Figure~\ref{fig:mcmc_no_bal_mc} we see that the model over-predicts the number of objects along the direction of largest deviation in $c_1$--$c_2$ space, whereas along the direction perpendicular (with the smallest deviation in $c_1$--$c_2$ space), the model under-predicts the number of objects.  This difference shows that our model favors changes in the slope over changes in curvature, and as a result under-estimates the amount of curvature in our data.  This is not surprising given that the average object has low curvature.  This difference could also be attributed to the underlying optical/UV SED not being well modeled by a powerlaw; see the Introduction and \citet[e.g.,][]{Schneider:2001,Shang:2005}.  In the bottom-right panel of Figure~\ref{fig:mcmc_no_bal_mc} the agreement for the BAL sample is better, but again the amount of curvature is under-estimated by our model.

\begin{figure*}[t]
\begin{center}
\includegraphics[width=6in]{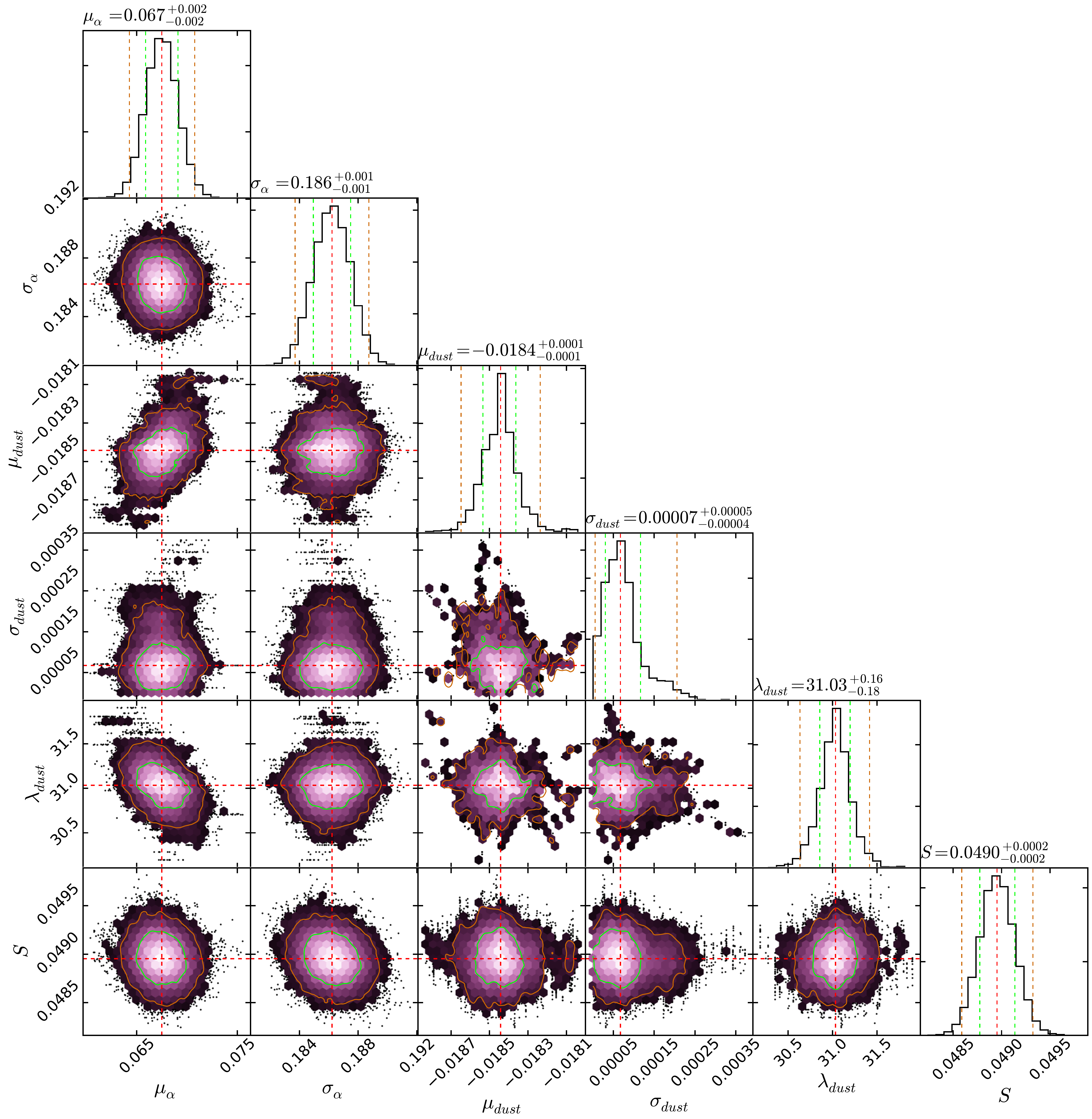}
\caption[Best fit population parameters for non-BAL quasars]{\label{fig:mcmc_no_bal_tri} The best fit population parameters for an SMC reddening law for the non-BAL quasars. The red dashed lines show the median for each variable, the 1$\sigma$ (2$\sigma$) confidence regions are shown in green (orange).  The $\dal$ distribution is fit with a normal distribution with mean $\mu_{\alpha}$ and width $\sigma_{\alpha}$. The $\ebv$ distribution is fit with a EMG with shape parameters $\mu_{\rm dust}$, $\sigma_{\rm dust}$, and $\lambda_{\rm dust}$ (see text). $S$ is the uncertainty associated with determining the $\mrel$.  
The mean values for these parameters are: $\{ \mu_{\alpha}, \sigma_{\alpha}, \mu_{\rm dust}$, $\sigma_{\rm dust} ,\lambda_{\rm dust}, S \}=\{ 0.067,0.186,-0.0184,7\times10^{-5},31.03,0.0490 \}$.
In the left column, third from the top, we see a positive correlation between the mean values of the $\dal$ and $\ebv$ distributions, showing there is a degeneracy between these variables.}
\end{center}
\end{figure*}

\begin{figure*}[t]
\begin{center}
\includegraphics[width=6in]{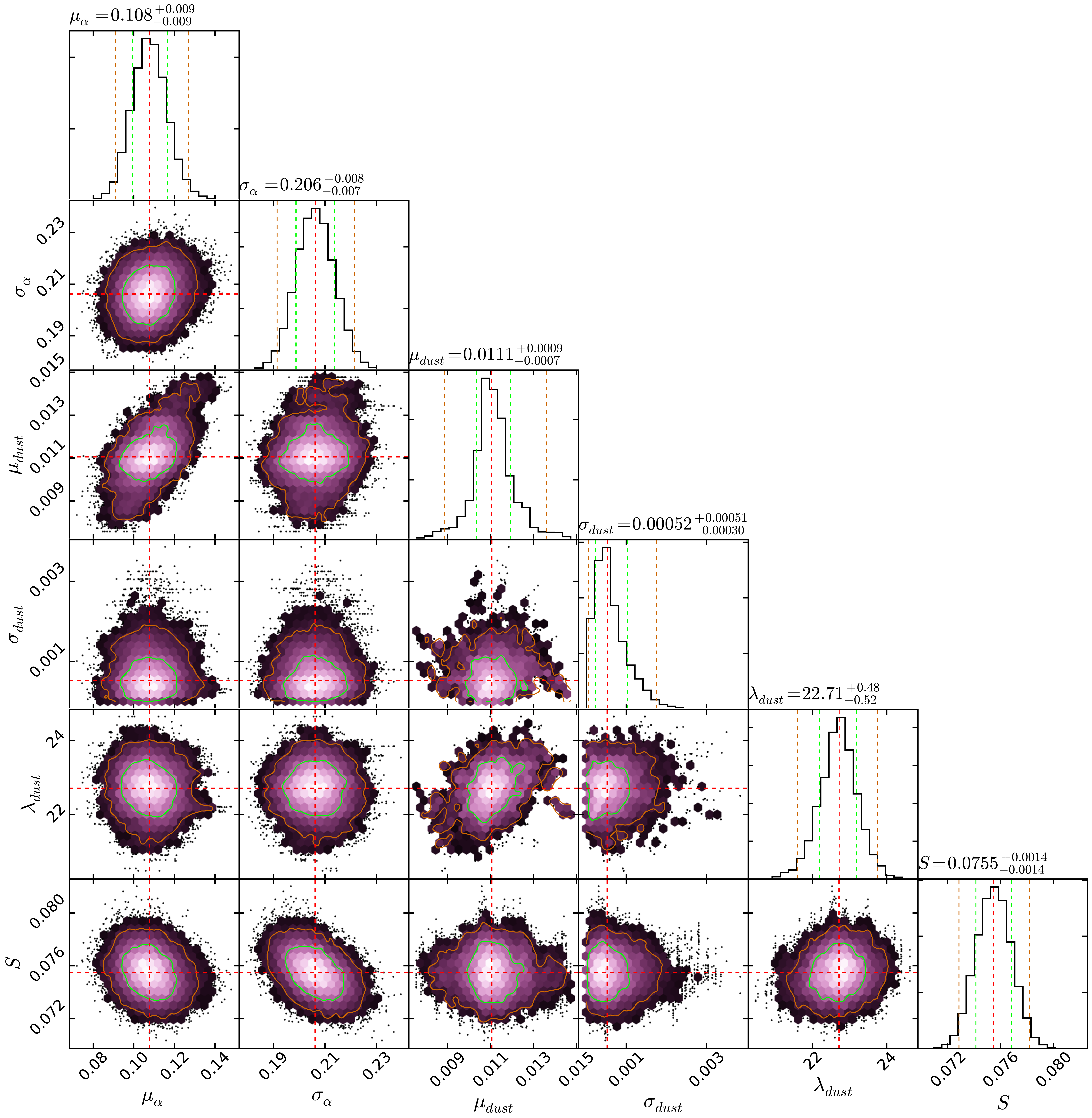}
\caption[Best fit population parameters for BAL quasars]{\label{fig:mcmc_bal_tri} Same as Figure~\ref{fig:mcmc_no_bal_tri} but for the BAL quasars. 
The mean values for these parameters are: $\{ \mu_{\alpha}, \sigma_{\alpha}, \mu_{\rm dust}$, $\sigma_{\rm dust} ,\lambda_{\rm dust}, S \}=\{ 0.108,0.206,0.0111,5.2\times10^{-4},22.71,0.0755 \}$.
As with the non-BAL sample we find a positive correlation between the mean values of the $\dal$ and $\ebv$ distributions, showing there is a degeneracy between these variables.  The BAL sample has a larger value for $\mu_\alpha$ and a smaller value for $\lambda_{\rm dust}$ than the non-BAL sample, indicating that the BAL quasars, on average, are intrinsically bluer within random error (see text) and have more dust reddening than the non-BAL sample.}
\end{center}
\end{figure*}

%The population parameters resulting from the fits to both reddening laws
 Figures~\ref{fig:mcmc_no_bal_tri} and~\ref{fig:mcmc_bal_tri} show the marginalized posterior distributions for the 1D and 2D projections for each of the six population parameters for the non-BAL and the BAL samples using the SMC reddening law. The resulting values of the population parameters for both reddening laws are given in Table~\ref{tab:hyp_fit_table} (see also the top and side panels of Figures~\ref{fig:indv_fits_SMC} and \ref{fig:indv_fits_L14}).

\begin{deluxetable*}{lrrrrr}
\tablewidth{0pt}

\tablecaption{\label{tab:hyp_fit_table} Hyperparameter Fit Values}
\tablehead{
\colhead{} & \multicolumn{2}{c}{SMC} & \colhead{} & \multicolumn{2}{c}{L14-QSO} \\
\cline{2-3} \cline{5-6} \\
\colhead{Population Parameter} & \colhead{non-BAL} & \colhead{BAL} & \colhead{} & \colhead{non-BAL} & \colhead{BAL}
}
\startdata
$\mu_{\alpha}$                                       & $0.067\pm0.002\phantom{0}$                 & $0.108^{+0.009\phantom{00}}_{-0.008}$      & & $0.039\pm0.002\phantom{0}$               & $0.021^{+0.008\phantom{00}}_{-0.007}$ \\
$\sigma_{\alpha}$                                  & $0.186\pm0.001\phantom{0}$                 & $0.206^{+0.008\phantom{00}}_{-0.007}$      & & $0.202\pm0.001\phantom{0}$               & $0.214\pm0.006$ \\
$\mu_{\rm dust}$                                   & $-0.0184\pm0.0001$                                  & $0.0111^{+0.0009\phantom{0}}_{-0.0007}$ & & $-0.0125\pm0.0001$                               & $0.0007^{+0.0004\phantom{0}}_{-0.0003}$\\
$\sigma_{\rm dust}$                              & $0.00007^{+0.00005}_{-0.00004}$         & $0.00052^{+0.00051}_{-00030}$                   & & $0.0004\pm0.0001$                                 & $0.00027^{+0.00023}_{-00015}$ \\
$\lambda_{\rm dust}\tablenotemark{a}$ & $31.03^{+0.16\phantom{000}}_{-0.18}$ & $22.71^{+0.48\phantom{000}}_{-0.52}$         & & $53.96^{+0.31\phantom{000}}_{-0.27}$ & $41.53^{+0.89\phantom{000}}_{-0.81}$\\
$S$                                                          & $0.0490\pm0.0002$                                    & $0.0755\pm0.0014$                                           & & $0.0469\pm0.0002$                                    & $0.0671\pm0.0012$
\enddata
\tablenotetext{a}{Smaller values indicate a heavier tail.}

\tablecomments{The 1$\sigma$ error regions are given for each value.}
\end{deluxetable*}

The projections shown in Figures~\ref{fig:mcmc_no_bal_tri} and~\ref{fig:mcmc_bal_tri} reveal information about the overall shapes of our $\dal$ and $\ebv$ distributions and correlations between the population parameters as characterized by the means and dispersions.  Both samples have a positive value for $\mu_{\alpha}$ indicating that the (intrinsic) mean spectral indices are bluer than the observed modal quasar ($\dal \gtrsim 0.07$).
Using Equation~(\ref{eqn:dal_2_anu}) to translate the mean $\dal$ values from Table~\ref{tab:hyp_fit_table} into observed spectral indices we find 
for the SMC law that the BAL sample is bluer ($\langle \alpha_\lambda \rangle = -1.83$) than the non-BAL sample ($\langle \alpha_\lambda \rangle = -1.79$).
%, and for the L14-QSO law the BAL sample is slightly redder ($\langle \alpha_\lambda \rangle = -1.74$) than the non-BAL sample ($\langle \alpha_\lambda \rangle = -1.76$).  
Although the values of $\mu_\alpha$ are different in terms of random error, after taking into account our systematic ability to measure the modal spectral index (related to $S$, see Appendix~\ref{sec:mcmc_model}), these values are well within 1$\sigma$ of each other (see the top panel of Figure~\ref{fig:indv_fits_SMC}). 
%This small difference indicates that the non-BAL and BAL quasar spectral indices are consistent with coming from the same distribution.
%statistically different between the two samples, these differences are within 1--2$\sigma$ of our (systematic) ability to measure the modal spectral index (related to $S$), indicating that the non-BAL and BAL quasars are consistent with having the same mean spectral index for both reddening laws.

In all cases $\sigma_{\rm dust}$ is consistent with $0$; since this parameter corresponds to the width of the Gaussian half of the EMG distribution, this means the distribution for $\ebv$ is consistent with coming from a pure exponential distribution that is offset from the origin by $\mu_{\rm dust}$.  We also find a positive correlation between $\mu_{\alpha}$ and $\mu_{\rm dust}$ (left column, third from the top in Figures~\ref{fig:mcmc_no_bal_tri} \& \ref{fig:mcmc_bal_tri}) showing there is a degeneracy between the mean values for $\dal$ and $\ebv$ in the model.
That is, it is difficult to distinguish between something that is intrinsically red and something that is dust reddened. 
The positive value of $\mu_{\rm dust}$ in the BAL sample suggests that all of the BAL quasars are consistent with having dust reddening, while the negative value on the non-BAL sample suggests that there are quasars ($\sim$60\%\footnote{The positive side of $\ebv$ is within 1$\sigma$ of 0}) with no dust reddening\footnote{If the modal quasar has a small amount of dust reddening, negative $\ebv$ values would indicate quasars with less dust than the mode.}.

Our population parameters for the SMC law are comparable to the shape parameters found in Table~2 of \citet{Hopkins:2004}. After converting the parameters to the same scale ($\sigma_{\alpha}^{\rm Hopkins} = \sigma_{\alpha}$, $\sigma_{\rm dust}^{\rm Hopkins} = 1/\lambda_{\rm dust}$, and $\sigma_{\rm Gaussian}^{\rm Hopkins}=\sigma_{\rm dust}$) we find $\sigma_{\alpha}$ (0.13 in Hopkins) to be higher for the non-BAL (0.18) and BAL (0.21) samples, indicating we see a larger range of spectral indices,
and $\lambda_{\rm dust}$ (22.22 in Hopkins) to be about the same for the BAL sample (22.71) and higher\footnote{Higher values mean fewer reddened quasars} for the non-BAL sample (31.03).  As mentioned above, our values for $\sigma_{\rm dust}$ are much smaller than the \citet{Hopkins:2004} value of 0.02.

\subsection{MCMC Results: Individual Parameters} \label{sec:mcmc_ip}

\begin{figure*}[t]
\begin{center}
\begin{tabular}{cc}
\includegraphics[width=3in]{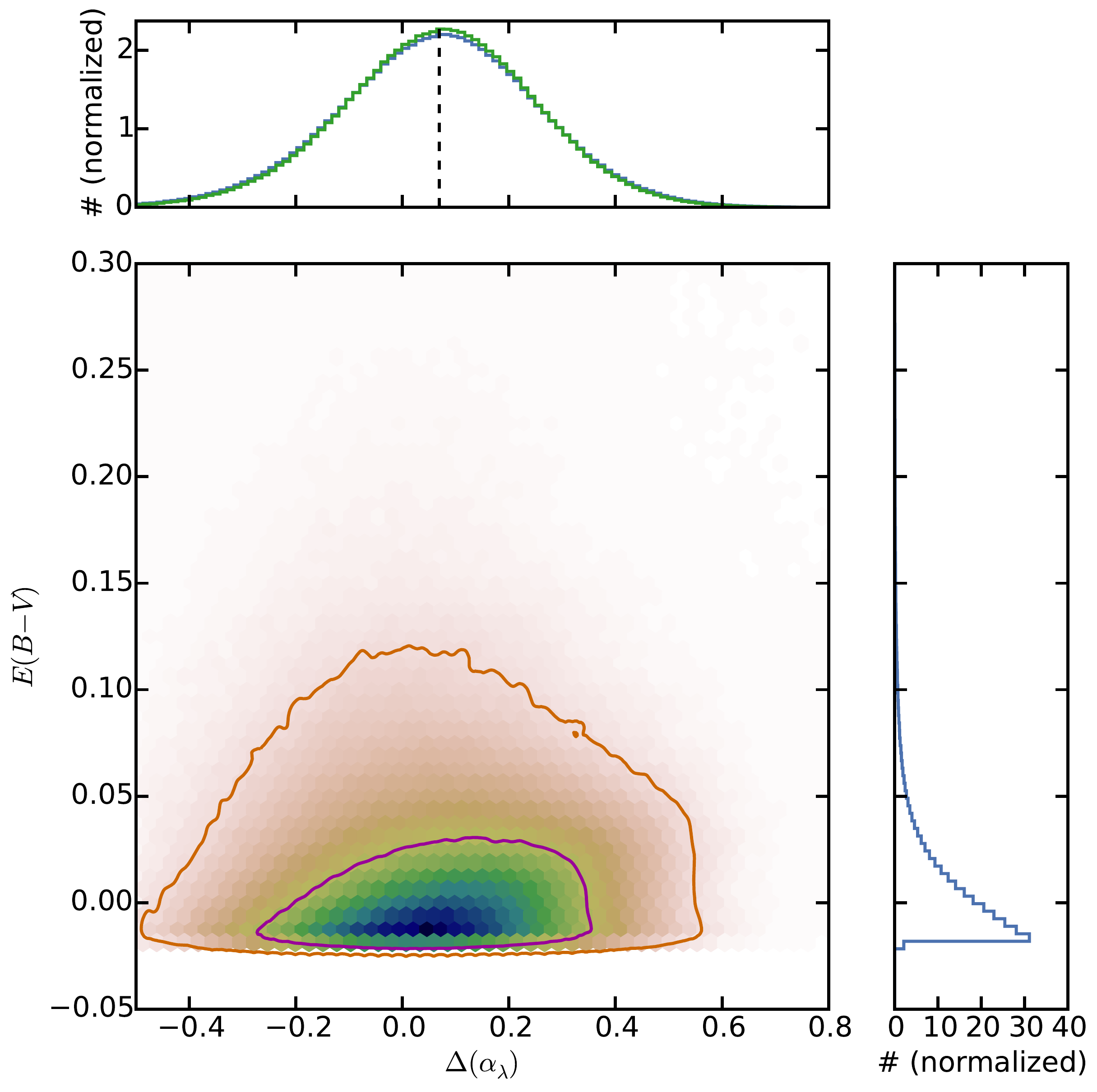} & \includegraphics[width=3in]{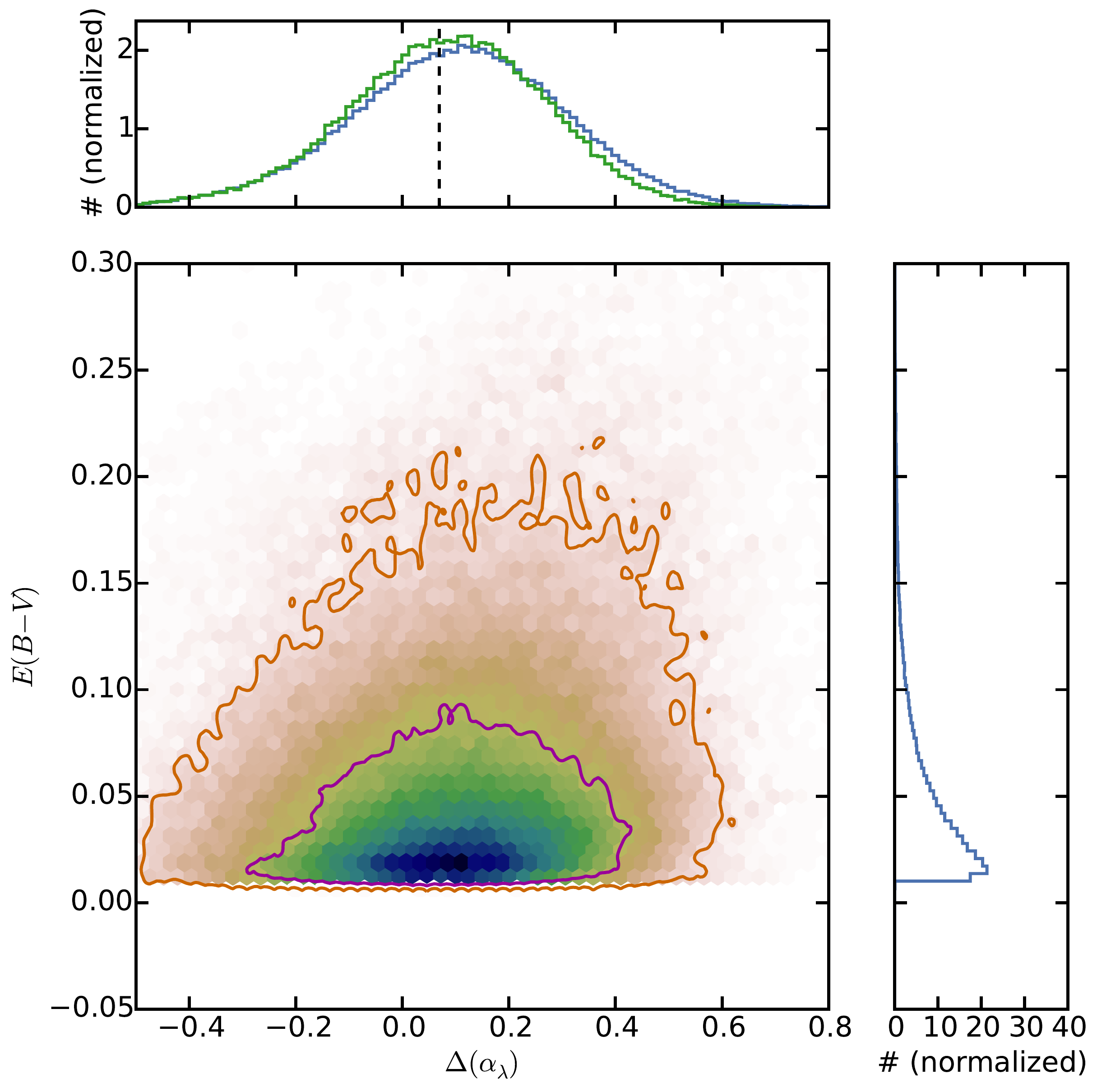} \\
\end{tabular}
\caption[Joint probability distribution of $\dal$ and $\ebv$]{\label{fig:indv_fits_SMC} Joint probability distribution of $\dal$ and $\ebv$ values for non-BAL ({\em left}) and BAL ({\em right}) quasars for the SMC reddening law. The 1$\sigma$ (2$\sigma$) contours are shown in purple (orange) and the fully marginalized distributions (blue) for each variable are shown as histograms.  Additionally the top panels show the marginalized $\dal$ distribution for the quasars with little to no reddening ($\ebv<0.04$) in green and the mean of the non-BAL sample shown as a vertical dashed line. The zero point on the x-axis represents the modal quasar that has $\alpha_{\lambda} =-1.72$ ($\alpha_{\nu}=-0.28$) and positive $\dal$ values are bluer.  The negative $\ebv$ values on the left argue that the mean non-BAL quasar is not dust reddened (and is $\sim$0.07 bluer in $\alpha_{\lambda}$ than we have assumed).  The positive $\ebv$ values on the right argue that, in contrast to non-BALs, BALs all have dust and are intrinsically bluer than the non-BALs by virtue of the centroid of the blue region. Moreover, at a given $\ebv$, the most probable value for $\dal$ is larger, suggesting that more heavily reddened BAL quasars are actually intrinsically bluer.}
\end{center}
\end{figure*}

\begin{figure*}[t]
\begin{center}
\begin{tabular}{cc}
\includegraphics[width=3in]{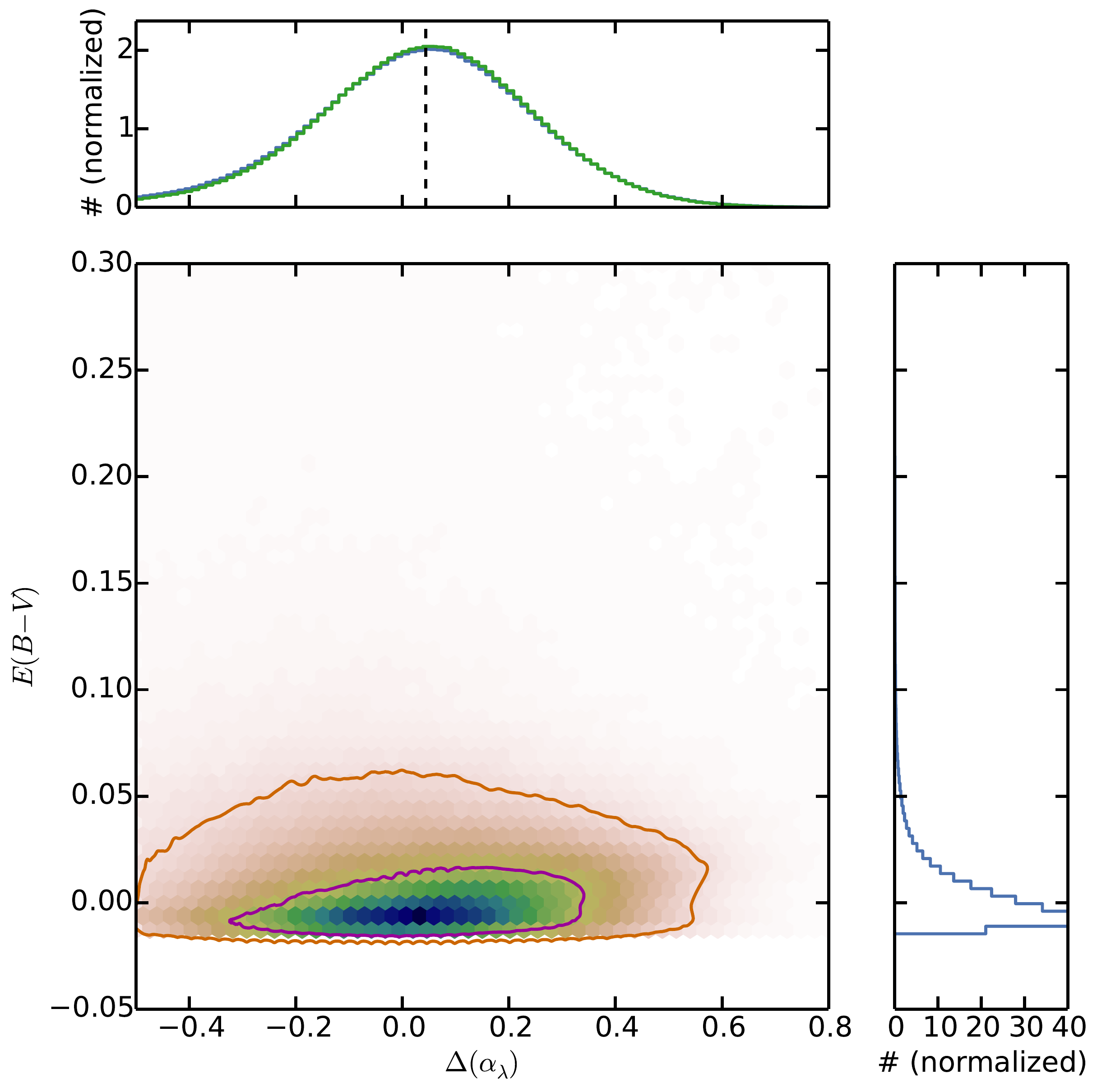} & \includegraphics[width=3in]{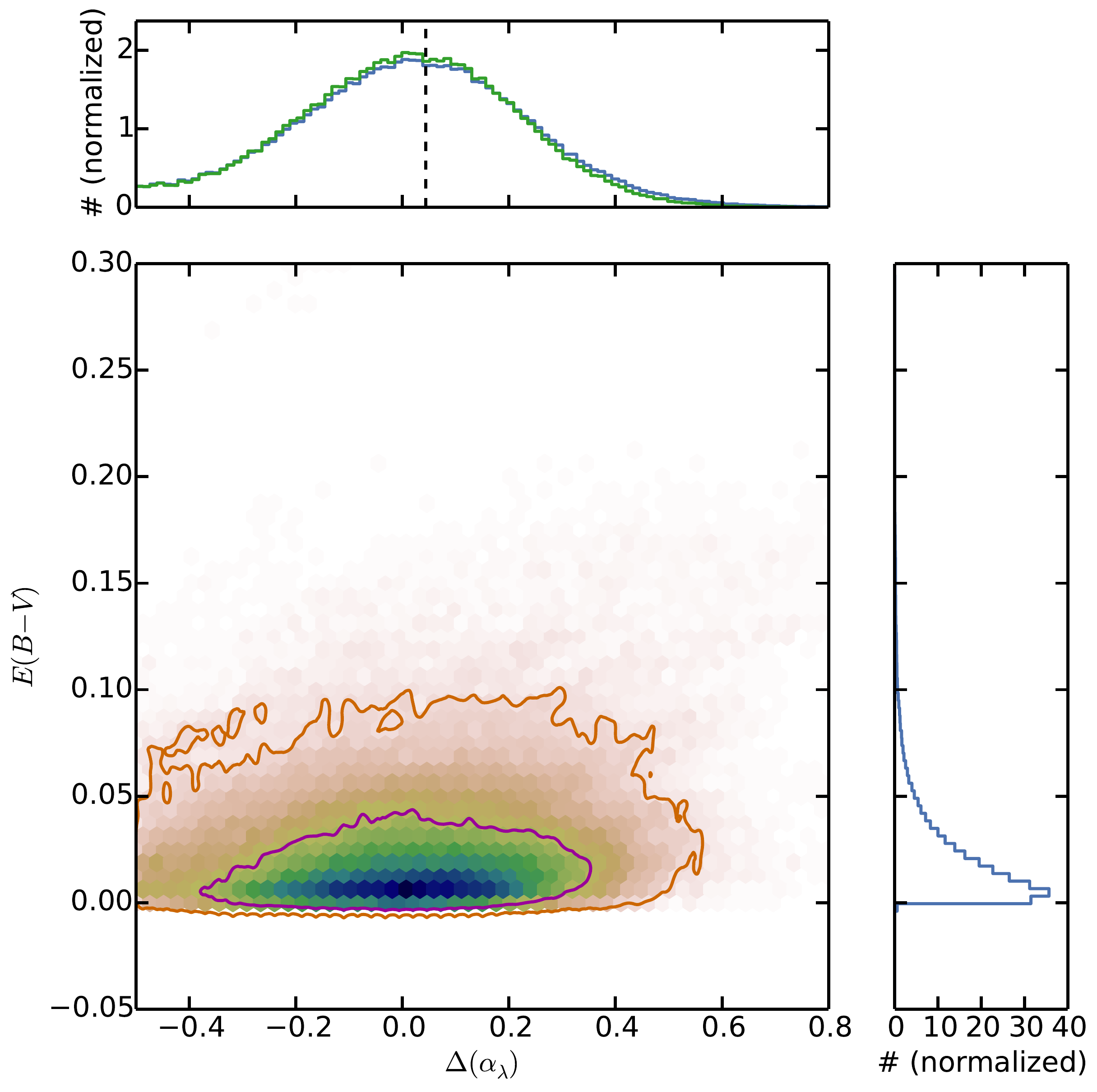} 
\end{tabular}
\caption[Joint probability distribution of $\dal$ and $\ebv$]{\label{fig:indv_fits_L14} Same as Figure~\ref{fig:indv_fits_SMC} but using the L14-QSO reddening law.  The general shapes of these distributions are the same but the $\ebv$ values are smaller, as expected for a steeper reddening law.  Again, the BAL quasars show that for a given $\ebv$, the most probable value for $\dal$ is larger, suggesting that the more heavily reddened BAL quasars are actually intrinsically bluer.}
\end{center}
\end{figure*}

Previous studies \citep[i.e.,][]{Reichard:2003a,Reichard:2003b} have avoided assigning physical meaning to their fit values for spectral index and amount of reddening due to the high degeneracy between these parameters \citep[e.g., Figure 3 of ][]{Reichard:2003a}. The correlation between the $\mu_{\alpha}$ and $\mu_{\rm dust}$ population parameters, and the correlations between the $\dal$ and $\ebv$ values for an individual quasar, are very similar to the degeneracies seen in previous work.  What makes our work different is the ability of our model to share this degeneracy across all levels of the model, from the individual fits up to the population parameters, in such a way that the degeneracy is broken in the marginalized distribution, as can be seen in Figures~\ref{fig:indv_fits_SMC} and \ref{fig:indv_fits_L14}. Put another way, the individual fits are regulated by the population parameters, kept within physical ranges by priors, and uncertainty is propagated through every level such that the marginalized distributions for $\dal$ and $\ebv$ follow the model we specified in Appendix~\ref{sec:mcmc_model}.
As a result the trends seen in Figures~\ref{fig:indv_fits_SMC} and \ref{fig:indv_fits_L14} are expected to be real trends in the data and are not caused by degeneracies in the model.  Table~\ref{tab:ind_fits} provides the $\alpha_{\lambda}$, $\ebv$, the covariance matrix ($C_{\alpha,\alpha}$,$C_{\alpha,\ebv}$, and $C_{\ebv,\ebv}$) between them, and the deviance information criterion (DIC, see Section~\ref{sec:DIC}) for each quasar and both reddening laws.

\begin{figure*}[t]
\begin{center}
\begin{tabular}{cc}
\includegraphics[width=3in]{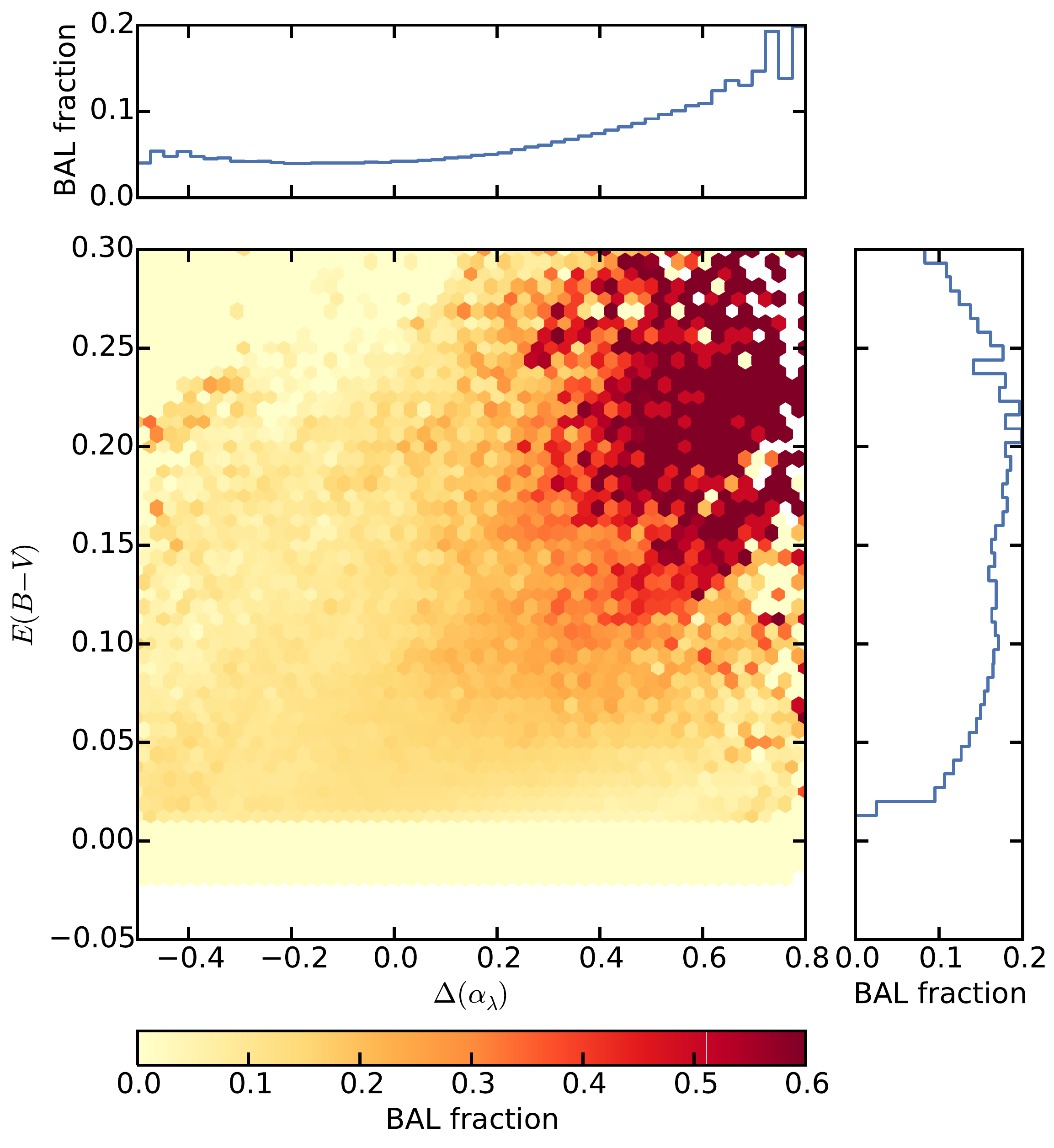} & \includegraphics[width=3in]{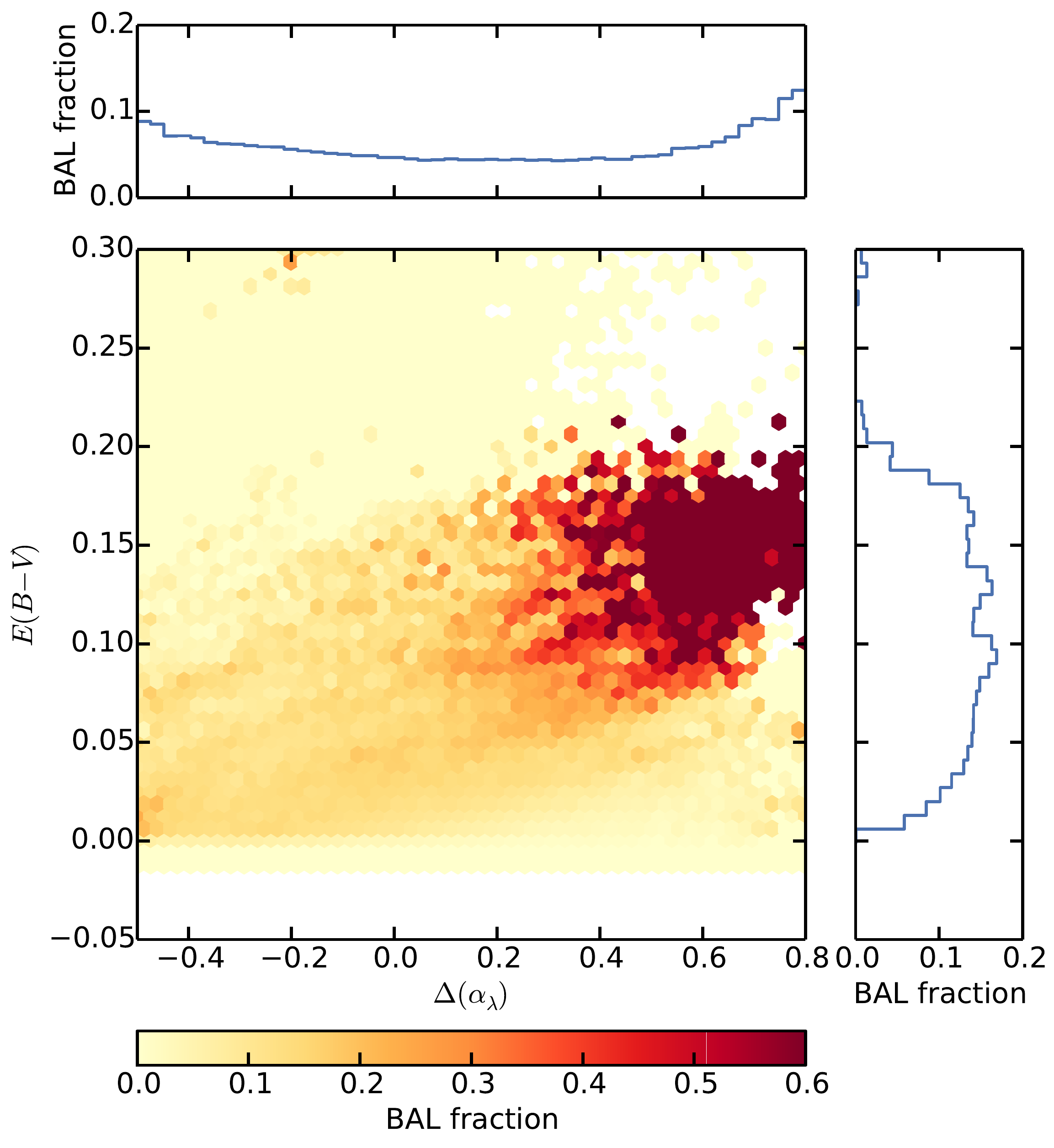} 
\end{tabular}
\caption[BAL fraction]{\label{fig:bal_frac} The fraction of BAL quasars as a function of $\dal$ and $\ebv$ for the SMC reddening law ({\em left}) and the L14-QSO reddening law ({\em right}).  The fully marginalized BAL fractions are shown in the histograms.  We can see the BAL fraction goes from $\sim0.01$ to $\sim0.6$ as the intrinsic color becomes bluer and the amount of dust increases.  Although the BAL fraction reaches $\sim0.6$ in the 2D parameters space, it does not exceed $\sim0.2$ after marginalizing over $\ebv$ or $\dal$.}
\end{center}
\end{figure*}

Figures~\ref{fig:indv_fits_SMC} and \ref{fig:indv_fits_L14} present the marginalized joint probability distributions for $\dal$ and $\ebv$ for each of our samples, i.e. the (2D) probability distribution for each {\em individual} quasar summed together and projected onto the $\dal$--$\ebv$ plane.  The top and side panels of each plot show the fully marginalized probability distributions for $\dal$ and $\ebv$ respectively; these distributions are the ones described by the population parameters in Table~\ref{tab:hyp_fit_table}.  The modal quasar ($\alpha_{\lambda} =-1.72$, $\alpha_{\nu}=-0.28$) represents the zero point of the x-axis and positive values of $\dal$ indicate quasars that are bluer than the mode.  

In Figure~\ref{fig:indv_fits_SMC} we show the results for fitting the SMC reddening law.  As was determined from the population parameters, we find that both the non-BAL and BAL samples are (on average) $\sim$0.07 bluer in $\alpha_{\lambda}$ than the modal quasar and the BAL sample shows more dust reddening.  The marginalized distributions for $\ebv$ argue that 2.5\% (13\%) of the $ugriz$ color-selected type 1 non-BAL (BAL) quasars in SDSS are consistent with $\ebv>0.1$ and 0.1\% (1.3\%) are consistent with $\ebv>0.2$.
Moreover, at a given $\ebv$ value, the BAL quasars are more likely to be bluer, suggesting that more heavily reddened BAL quasars are actually intrinsically bluer.  This trend is not seen in the non-BAL sample.  In both plots we see a lack of quasars that are intrinsically red and heavily dust reddened.  This deficit could be caused by an observational bias, i.e. quasars in this region have colors that are missed by the selection algorithm.  If this is not the case, then this might suggest that the heaviest levels of extinction are associated with the quasar itself rather than the host galaxy, otherwise there would be no link between the intrinsic color and the reddening (unless there is also a link between the host galaxy and the intrinsic quasar color).

\renewcommand{\tabcolsep}{.5mm}
\renewcommand{\arraystretch}{1.5}
\afterpage{
\begin{turnpage}
\begin{deluxetable*}{lrrcccccrrccccc}
\tabletypesize{\scriptsize}
%\rotate
\tablewidth{0pc}
\tablecolumns{15}
\tablecaption{Individual Fit Values\label{tab:ind_fits}}
\tablehead{
\colhead{} & \multicolumn{6}{c}{SMC} & \colhead{} & \multicolumn{6}{c}{L14-QSO} & \colhead{} \\
\cline{2-7} \cline{9-14} \\
 \colhead{Name (J2000)} & \colhead{$\dal$} & \colhead{$\ebv$} & \colhead{$C_{\alpha,\alpha}$} & \colhead{$C_{\ebv,\ebv}$} & \colhead{$C_{\alpha,\ebv}$} & \colhead{DIC} & \colhead{} & \colhead{$\dal$} & \colhead{$\ebv$} & \colhead{$C_{\alpha,\alpha}$} & \colhead{$C_{\ebv,\ebv}$} & \colhead{$C_{\alpha,\ebv}$} & \colhead{DIC} & \colhead{BAL}  }
\startdata
095628.84+461000.7 & $-0.04^{+0.19}_{-0.19}$ & $0.114^{+0.023}_{-0.023}$ & $2.22\times10^{-1}$ & $5.40\times10^{-4}$ & $9.36\times10^{-3}$ & -27.04 &  & $-0.17^{+0.19}_{-0.17}$ & $0.054^{+0.012}_{-0.011}$ & $2.22\times10^{-1}$ & $5.40\times10^{-4}$ & $9.36\times10^{-3}$ & -20.94 & 1\\
102630.35+424632.0 & $0.11^{+0.12}_{-0.10}$ & $-0.007^{+0.014}_{-0.008}$ & $8.21\times10^{-2}$ & $1.32\times10^{-4}$ & $2.49\times10^{-3}$ & -26.98 &  & $0.08^{+0.13}_{-0.11}$ & $-0.006^{+0.008}_{-0.005}$ & $8.21\times10^{-2}$ & $1.32\times10^{-4}$ & $2.49\times10^{-3}$ & -18.34 & 0\\
104543.39+533203.0 & $-0.05^{+0.17}_{-0.15}$ & $0.021^{+0.025}_{-0.022}$ & $1.55\times10^{-1}$ & $5.04\times10^{-4}$ & $7.71\times10^{-3}$ & -25.85 &  & $-0.09^{+0.17}_{-0.16}$ & $0.010^{+0.013}_{-0.012}$ & $1.55\times10^{-1}$ & $5.04\times10^{-4}$ & $7.71\times10^{-3}$ & -21.47 & 0\\
105105.81+274934.4 & $-0.04^{+0.16}_{-0.16}$ & $0.087^{+0.029}_{-0.030}$ & $1.60\times10^{-1}$ & $8.92\times10^{-4}$ & $1.08\times10^{-2}$ & -29.52 &  & $-0.10^{+0.15}_{-0.15}$ & $0.047^{+0.017}_{-0.017}$ & $1.60\times10^{-1}$ & $8.92\times10^{-4}$ & $1.08\times10^{-2}$ & -12.32 & 0\\
120540.29+384636.9 & $0.21^{+0.09}_{-0.09}$ & $-0.007^{+0.017}_{-0.008}$ & $5.05\times10^{-2}$ & $2.00\times10^{-4}$ & $1.77\times10^{-3}$ & -31.00 &  & $0.20^{+0.08}_{-0.08}$ & $-0.006^{+0.010}_{-0.005}$ & $5.05\times10^{-2}$ & $2.00\times10^{-4}$ & $1.77\times10^{-3}$ & -13.69 & 0\\
132655.12+254811.2 & $0.23^{+0.09}_{-0.08}$ & $0.020^{+0.013}_{-0.007}$ & $4.84\times10^{-2}$ & $1.22\times10^{-4}$ & $1.58\times10^{-3}$ & -21.31 &  & $0.15^{+0.08}_{-0.07}$ & $0.005^{+0.006}_{-0.003}$ & $4.84\times10^{-2}$ & $1.22\times10^{-4}$ & $1.58\times10^{-3}$ & -17.48 & 1\\
143047.61+304532.2 & $0.11^{+0.18}_{-0.16}$ & $0.039^{+0.020}_{-0.017}$ & $1.78\times10^{-1}$ & $3.18\times10^{-4}$ & $6.06\times10^{-3}$ & -25.78 &  & $0.06^{+0.19}_{-0.17}$ & $0.017^{+0.011}_{-0.009}$ & $1.78\times10^{-1}$ & $3.18\times10^{-4}$ & $6.06\times10^{-3}$ & -17.55 & 1\\
144610.88+132815.2 & $0.10^{+0.17}_{-0.18}$ & $0.110^{+0.033}_{-0.036}$ & $1.94\times10^{-1}$ & $1.19\times10^{-3}$ & $1.41\times10^{-2}$ & -20.79 &  & $-0.02^{+0.14}_{-0.14}$ & $0.057^{+0.016}_{-0.018}$ & $1.94\times10^{-1}$ & $1.19\times10^{-3}$ & $1.41\times10^{-2}$ & -18.56 & 1\\
154117.05+100910.9 & $0.25^{+0.10}_{-0.08}$ & $0.003^{+0.017}_{-0.014}$ & $4.91\times10^{-2}$ & $2.18\times10^{-4}$ & $2.85\times10^{-3}$ & -33.60 &  & $0.26^{+0.08}_{-0.08}$ & $0.003^{+0.008}_{-0.008}$ & $4.91\times10^{-2}$ & $2.18\times10^{-4}$ & $2.85\times10^{-3}$ & -21.24 & 0\\
155956.92+204522.0 & $0.00^{+0.20}_{-0.19}$ & $0.059^{+0.021}_{-0.020}$ & $2.34\times10^{-1}$ & $4.15\times10^{-4}$ & $8.21\times10^{-3}$ & -27.46 &  & $-0.10^{+0.19}_{-0.19}$ & $0.025^{+0.010}_{-0.009}$ & $2.34\times10^{-1}$ & $4.15\times10^{-4}$ & $8.21\times10^{-3}$ & -12.94 & 1\\
\enddata
\tablecomments{This table is available in its entirety in a machine-readable form in the electronic version of this journal. A portion is shown here for guidance regarding its form and content.}
\end{deluxetable*}

\end{turnpage}
}
\renewcommand{\arraystretch}{1}

In Figure~\ref{fig:indv_fits_L14} we see qualitatively the same trends for the L14-QSO reddening law as we did for the SMC law, the biggest differences being the scaling of the y-axis and a shift in the $\dal$ to smaller values. Here the marginalized distributions for $\ebv$ argue that 0.2\% (1.7\%) of the non-BAL (BAL) quasars are consistent with $\ebv>0.1$ and 0.002\% (0.022\%) are consistent with $\ebv>0.2$.

Using this data set, we can also look at the observed BAL fraction as a function of $\dal$ and $\ebv$.  We have made no attempt to find the intrinsic BAL fraction \citep[e.g., see][]{Hewett:2003, Allen:2011}, instead, we are looking at how the relative fraction of observed BALs changes as a function of these parameters.  Figure~\ref{fig:bal_frac} shows the BAL fraction for the SMC and L14-QSO reddening laws.  As with the previous plots, we see that the changes in the BAL fraction are qualitatively similar for both reddening laws: it increases from $\sim0.01$ for red quasars with little dust reddening, up to $\sim0.6$ for blue quasars with heavy dust reddening.  After marginalizing over $\ebv$ we see that the BAL fraction ranges between $\sim0.05$ and $\sim0.2$ and it clearly increases as quasars become bluer.   This trend has only become clear due to our larger data set, and is in comparison to previous studies \citep[e.g. ][]{Tolea:2002,Hewett:2003,Reichard:2003b} that were only able to measure the mean BAL fraction of $\sim20\%$.
%This is in contrast to the constant BAL fraction of $\sim20\%$ seen throughout the literature \citep[e.g. ][]{Tolea:2002,Hewett:2003,Reichard:2003b}.  
As with Figures~\ref{fig:indv_fits_SMC} and \ref{fig:indv_fits_L14} we have marginalized over the probability distributions for each individual quasar, meaning the trend seen here is expected to be real (up to any observational biases) and not an artifact of our model. 

This analysis suggests that the parent populations for BALs and non-BALs are slightly different (BALs being intrinsically bluer, on average).  However, we need to be careful that selection effects are not involved \citep[e.g., see][]{Reichard:2003b, Allen:2011}.

%Although we find the parent distributions for our BAL and non-BAL samples to be different, it can be argued  that the parent distributions are not necessarily intrinsically different. The observed difference could be caused by a selection effect. In color-color space intrinsically blue quasars overlap the stellar locus between the redshifts of 2.7 and 3 [are these numbers right], making them difficult to identify.  If these quasars have a sufficient amount of dust [should figure out how much] they will be shifted out of the stellar locus making them easier to identify.  Since BAL quasars are more heavily reddened than their non-BAL counterparts, we are more likely to identify intrinsically blue BALs than intrinsically blue non-BALs in this redshift range.  Alternatively the parent distributions could be intrinsically different and bluer quasars have a large probability of have BALs (e.g. bluer quasars are more likely to have a disk wind) than redder ones.

\subsection{Which Reddening Law Fits Better?}\label{sec:DIC}

Given that we used two different reddening laws to fit our data, it is natural to ask: which reddening law fits the data better?  To answer this question we use the DIC.  The DIC is a generalization of the Bayesian information criterion (BIC) such that it can be directly applied to the output of a Markov Chain Monte Carlo (MCMC) fit, and can be used to directly compare multiple models fit to the same data.  The DIC is made up of two terms, one that is related to the likelihood of the model explaining the data (similar to $\chi^2$), and a penalty term based on the complexity of the model (i.e. the effective degrees of freedom after taking into account the covariances between the fitted variables).  

Models that have smaller DIC values fit the data better than models with higher DIC values, and the size of this difference gives a measure of how much better the fit is.  When the difference (DIC$_{\rm model 1}-$DIC$_{\rm model 2}$) is less than 2 both models fit the data well, when it is between 2 and 6 there is positive evidence for the lower model fitting better, and when it is greater than 6 there is strong evidence \citep{Kass:1995}.  We have calculated the DIC for each individual quasar and each reddening law to see what model fits better, these values are included in Table~\ref{tab:ind_fits}.  Table~\ref{tab:model_compare} summarizes the comparison of the two models.  Overall we find the SMC reddening law does a better job fitting our data, and to the end of applying the best correction in the mean sense, it will be used exclusively for the remainder of this articale.  That said, it should be noted that, while the SMC law is the best fit for our sample, there are a few objects for which other investigators have found that MW dust is a better fit.  For a recent example (four quasars), see \citet{Capellupo:2015}.  We do find that up to 4\% (11\%) of non-BAL (BAL) quasars show signs of contributions from multiple scattering; for future work it may be interesting to consider this subset to see how they differ from the main sample.  

%for future work it may be interesting to consider the subset of quasars that are better fit by the multiple-scattering law, and see how they differ from the main sample.

%[However, for future work it may be important to consider that a multiple-scattering law better fits a fraction of the data (and is most consistent with the dashed line in Figure 4.4 -- what do you make of that?]

\begin{deluxetable}{lrrrrr}
\tablewidth{0pt}

\tablecaption{\label{tab:model_compare} Reddening Law Comparison}
\tablehead{
\colhead{} & \multicolumn{2}{c}{non-BAL} & \colhead{} & \multicolumn{2}{c}{BAL} \\
\cline{2-3} \cline{5-6}\\
\colhead{Evidence} & \colhead{Number} & \colhead{Fraction} & \colhead{} & \colhead{Number} & \colhead{Fraction}
}
\startdata
SMC: strong            & 23443 & 68.5\% & & 1131 & 64.9\% \\
SMC: positive          & 9291 & 27.1\% & & 414 & 23.7\% \\
Both                          & 1470 & 4.3\% & & 192 & 11.0\% \\
L14-QSO: positive & 16 & 0.05\% & & 5 & 0.3\% \\
L14-QSO: strong    & 6 & 0.01\% & & 2 & 0.1\%
\enddata
\end{deluxetable}

\section{Composite Spectra} \label{sec:spectra}
\begin{figure*}
\begin{center}
\includegraphics[width=6in]{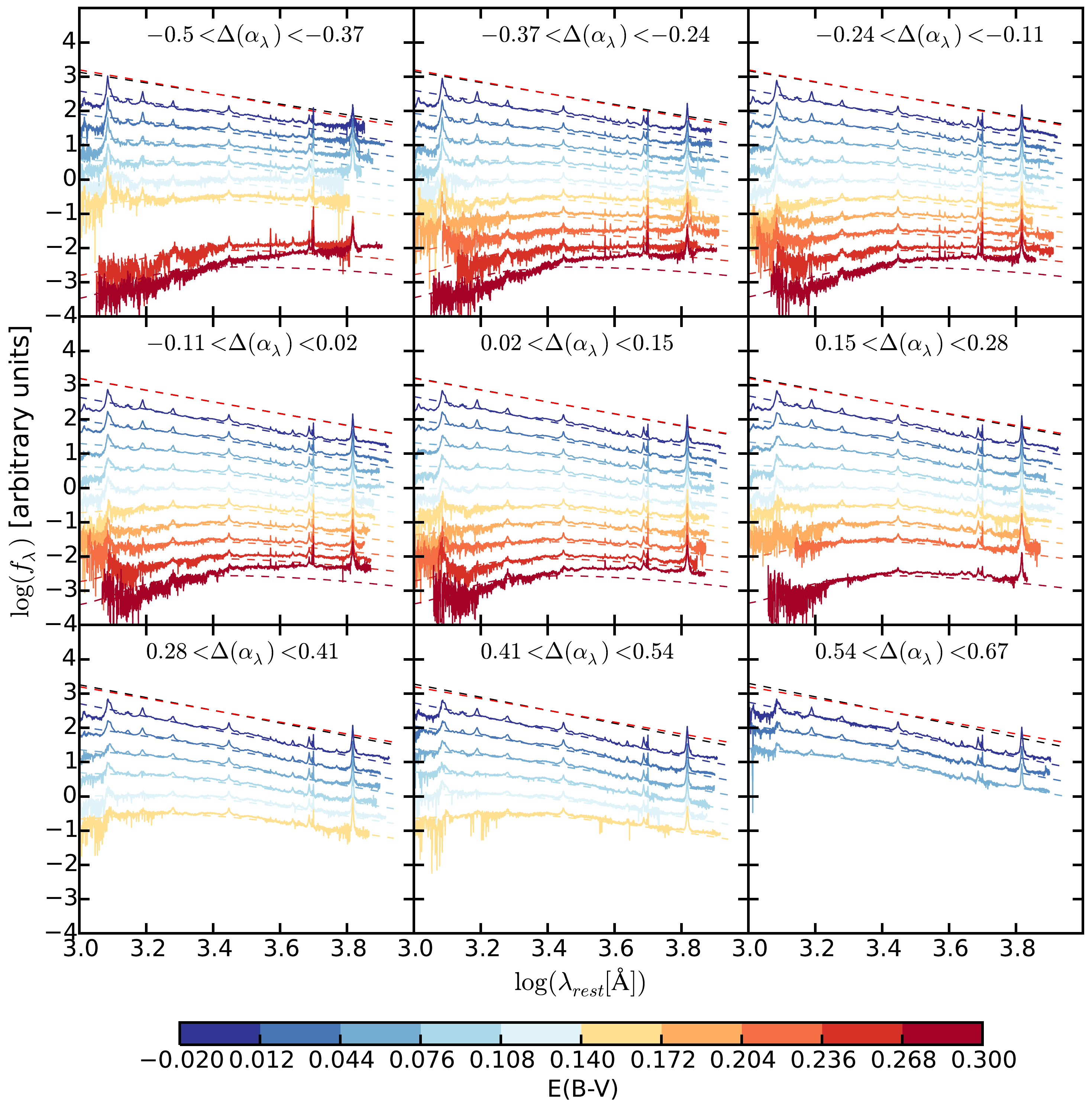}
\caption[Composite spectra for non-BAL quasars]{\label{fig:spec} Composite spectra for non-BAL quasars (solid lines) for nine evenly spaced bins along $\dal$ (subplots going intrinsically red to blue from the top left to the bottom right) and for 10 evenly spaced bins along $\ebv$ (various colors) based on the fits to the SMC reddening law.  The black dashed line in each panel indicates the typical powerlaw for each quasar in that $\dal$ bin and the red dashed line shows the observed modal powerlaw ($\dal=0$) for comparison.  The colored dashed lines show the typical powerlaw reddened by the $\ebv$ value corresponding to each color. The middle-left panel shows objects with spectral indices most similar to the modal powerlaw.  The objects best fit by an intrinsically bluer spectrum demonstrate that such objects exist, but the bluest spectra that are best fit without any dust are not perfect fits.  This could be a sign of mis-fitting, or may be an indication of the spectral turn-over within the fitting region (which is expected from \citet{Shang:2005}).  The most reddened spectra in each panel need a steeper reddening law to explain their curvature.  Each composite spectrum is largely in agreement with the bin values found from fitting the photometry. The data used to create this figure are available.}
\end{center}
\end{figure*}

\begin{figure*}
\begin{center}
\includegraphics[width=6in]{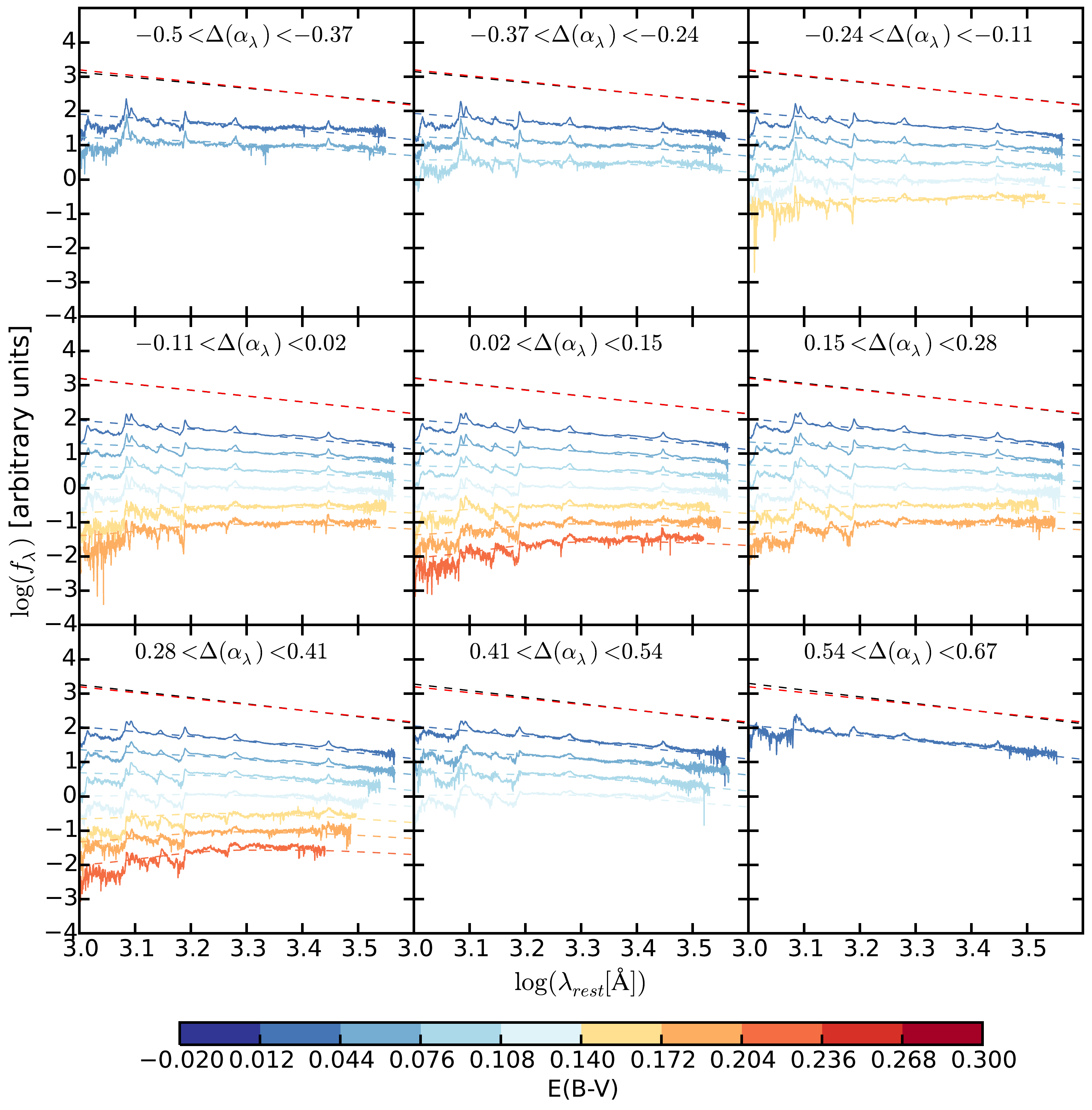}
\caption[Composite spectra for BAL quasars]{\label{fig:bal_spec} 
%Composite spectra for BAL quasars (solid lines) for ten evenly spaced bins along $\ebv$ (various colors) and nine evenly spaced bins along 2.5$\dal$ (subplots).  The black dashed line in each panel indicates the typical powerlaw for each quasar in that 2.5$\dal$ bin and the red dashed line shows the observed modal powerlaw ($\dal=0$) for comparison.  The colored dashed lines show the typical powerlaw reddened by the $\ebv$ value corresponding to each color. The middle-left panel shows objects with spectral indices most similar to the modal powerlaw.  
Same as Figure~\ref{fig:spec} but for the BAL sample (subplots going intrinsically red to blue from the top left to the bottom right).  Within each panel there is an increase in the depth of the \ion{C}{4} absorption trough ($\log{(\lambda)}\sim3.2$) as the $\ebv$ values increase.  The \ion{C}{4} absorption trough also becomes wider as quasars become bluer (e.g. the trough extends out to the nearby \ion{Si}{4} line for the bluest quasars).
%in the fraction of LoBAL quasars (absorption troughs blueward of \ion{Mg}{2}) as the $\ebv$ values increase. This trend is expected from the orientation-dependent picture presented in \citet{Weymann:1991}. We also see that every spectrum is consistent with at least some amount of dust reddening ($\ebv>0.012$). 
Each composite spectrum is largely in agreement with the bin values found from fitting the photometry. The data used to create this figure are available.}
\end{center}
\end{figure*}

\citet{Hopkins:2004} discuss composite spectra as functions of the ($c_1$,$c_2$) parameter space.  Here we construct composite spectra based on the more physical parameter space of $\dal$ and $\ebv$ from the previous section.  Our composites are constructed in the same way as the \citet{Vanden-Berk:2001} quasar composite, using similar code. In brief, the quasars are sorted by redshift, shifted to their rest frames, re-sampled onto a common 1\AA\ grid, and normalized by the overlap with the current composite.  For our composite spectra we used the improved SDSS redshifts cataloged in \citet{Hewett:2010}.  To preserve the continuum shape in the composite we chose to use a weighted geometric mean to combine the spectra \citep{Vanden-Berk:2001,Reichard:2003b}.

The spectra were split into 10 equally spaced bins in both $\dal$ and $\ebv$. The MCMC sampler from Section~\ref{sec:mcmc} produced full probability distributions for each quasar in this parameter space, allowing us to find the probability for each quasar belonging to any given bin.
This probability was used as the weight when calculating the geometric mean in each bin. To prevent quasars with high variance (i.e. spread across many bins) from contaminating the sample, only quasars with a probability of being in a bin greater than 10\% were used. After this cut, composites were made for any bin with more than 10 quasars.

\subsection{Red vs.\ Reddened Spectra}

Figures~\ref{fig:spec} and~\ref{fig:bal_spec} show the resulting composites for the non-BAL and BAL samples respectively and the flux values are provided. Each panel in these figures corresponds to a single $\dal$ bin (the \nth{10} bin for each sample had too few quasars to stack) and each color corresponds to a single $\ebv$ bin. The black dashed lines show the typical intrinsic powerlaw in each bin (red to blue going from the top left to bottom right) with the red dashed lines showing the global modal powerlaw, $\alpha_{\lambda}=-1.72$ ($\alpha_{\nu}=-0.28$), for comparison.  The color dashed lines show the typical powerlaw for each bin reddened by the $\ebv$ values corresponding to each color.  In most cases we see that the color dashed lines are in agreement with the stacked spectra in each bin, indicating that the $\dal$ and $\ebv$ values found in the previous section are accurate.  The notable exceptions are the most reddened bin for almost all the $\dal$ bins, these spectra tend to show more curvature than would be expected from the fits to their photometry suggesting, that, if anything, those quasars are better fit by a steeper reddening law (such as multiple-scattering). 

We also see slight deviations in the bluest bin for the least reddened spectra.  One potential cause for this deviation is if the underlying continuum between 1\,$\mu$m and 1216\,\AA\ is not well described by a power law.  
As noted in the introduction, there are a number of reasons to expect such deviations from a power-law continuum.

%for an idealized black body treatment of an thin accretion disk, such curvature can come about from sampling too close to the frequency/wavelength of the black body peak, where the effective slope is a function of mass, accretion rate, spin, and orientation.   More complex treatment of accretion disk SEDs \citep[e.g.][]{Hubeny:2000} also leads to deviations from power-law continua even well away from the UV peak of the SED.

\iffalse
The continuum emission from a quasar is thought to come from a modified blackbody \citep{Shakura:1976} that has been seen to turn over at wavelengths of about $\sim$1000--1300\,\AA\ \citep{Shang:2005}. If this turn over happens within the fitting range of our model, these quasars will be erroneously fit with large $\ebv$ values and have curvatures that are different than that caused by an SMC law. %\citet{Krawczyk:2013} found that more luminous quasars tended to turn over at longer wavelengths than less luminous quasars, if these objects are also intrinsically bluer, 
\fi

For the BAL quasars in Figure~\ref{fig:bal_spec} we see a trend in \ion{C}{4} absorption ($\log{(\lambda)}\sim3.2$) with $\dal$ and $\ebv$.  At a given $\dal$ value the absorption trough for \ion{C}{4} deepens as $\ebv$ increases.  When holding $\ebv$ at a fixed value, the absorption trough becomes wider (and extends to higher velocities) 
for bluer colors than redder colors.  These trends are consistent with \citet{Baskin:2013} who found \ion{C}{4} troughs become wider as the strength of \ion{He}{2} decreased, a trend we find also corresponds to the non-BAL quasars being bluer (see Section~\ref{sec:indv_lines}).  They also found the depth of the \ion{C}{4} trough increased as $\alpha_{\rm UVl}$ (measured between 1710 and 3000\,\AA) became redder, a quantity they associated with dust reddening.

\subsection{Individual Lines} \label{sec:indv_lines}
\begin{figure*}[t]
\begin{center}
\includegraphics[width=6in]{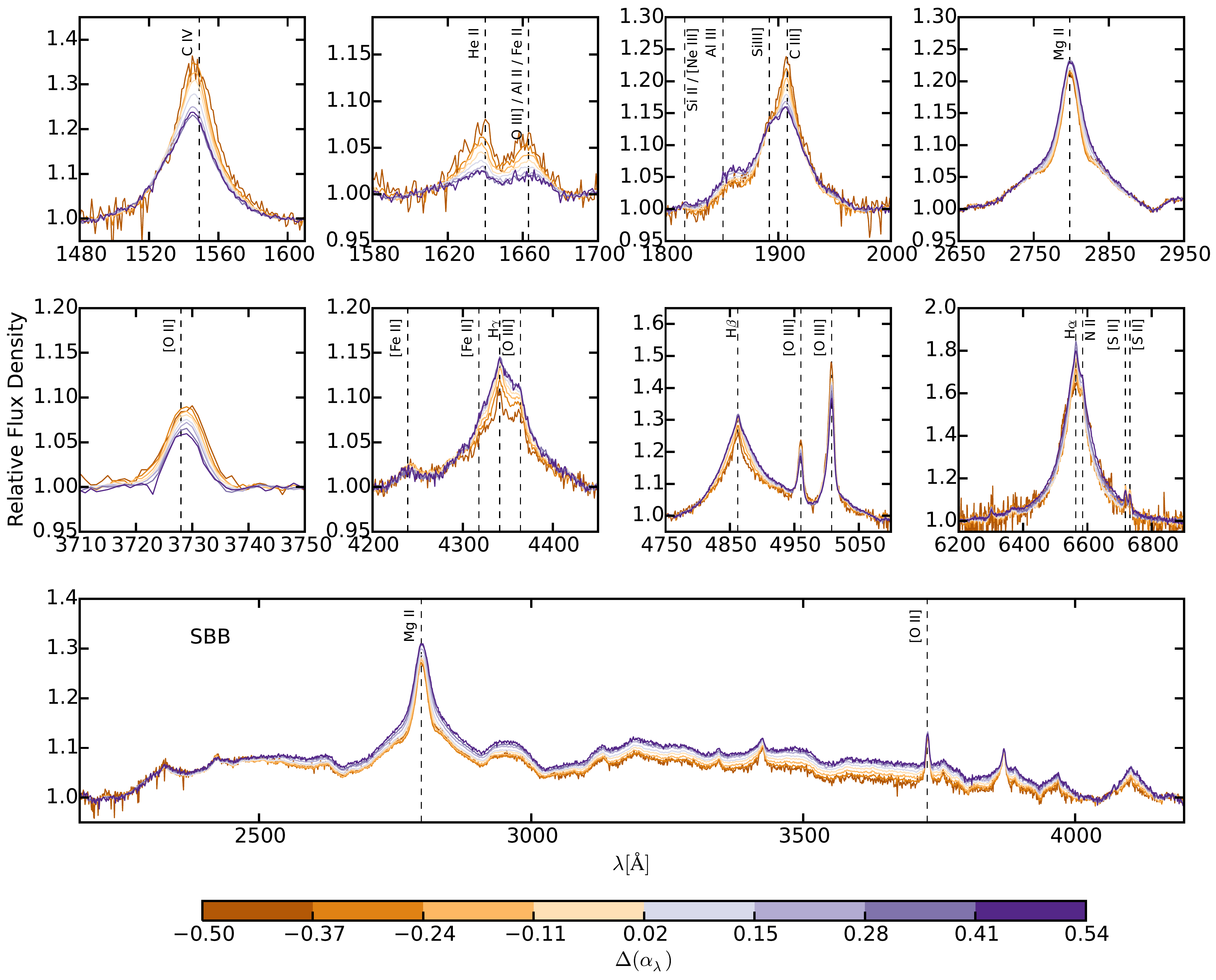}
\caption[Expanded emission-line regions for non-BAL quasars]{\label{fig:ind_spec} Expanded emission-line regions from the lowest $\ebv$ bin from Figure~\ref{fig:spec} ($\ebv<0.012$). ``SBB'' refers to the 3000 \AA\ (or small blue) bump.  The colors represent the $\dal$ value for each spectrum.  The spectra are normalized such that their continua agree at the edges of the panels.}
\end{center}
\end{figure*}

\citet{Richards:2003} constructed several composite spectra based on the observed $\dgi$ relative colors. They were able to construct four spectra that had little dust reddening and span different spectral indices.  By comparing these spectra they found several trends in individual spectral lines related to the intrinsic slopes. Our analysis provides a more robust metric for distinguish intrinsically red and dust reddened quasars, allowing us to create a cleaner sample of quasars that are thought to be relatively free of dust. This process is useful as one expects a physical link between the continuum and the (photoionized) gas in the BELR.

\subsubsection{Trends with Spectral Index}

In this section we explore the trends in some individual emission lines as a function of the intrinsic spectral index for the non-BAL quasars consistent with little to no reddening (i.e. the lowest $\ebv$ bin).  To prevent being influenced by known trends due to quasar luminosity, we picked sub-samples from each bin such that the distribution of \ltwofive, taken from \citet{Krawczyk:2013}, was the same for every sample. When doing this we excluded the two bluest bins from the analysis seeing as their \ltwofive\ distributions were significantly lower than the other bins. 
%\footnote{This lower \ltwofive\ distribution is cased by lower redshift distribution for the bluest quasars (see Figure~\ref{l_z_alpha}).}.  
The resulting luminosity distribution for the non-BAL sample had $44.4<\log{(L_{2500\,\AA}{\rm \ [ergs\ s^{-1}]})}<47$ and a median of 45.7, while the BAL sample had $46<\log{(L_{2500\,\AA}{\rm \ [ergs\ s^{-1}]})}<46.7$ and a median of 46.2.

The major emission line regions are shown in Figure~\ref{fig:ind_spec}. These spectra have been normalized such that they are unity at the edges of each plot\footnote{This is achieved by dividing out the powerlaw continuum connecting the spectra at the edges of each plot}. Below we comment on each of the the emission lines. %[make a note that the abstract for \citet{Richards:2003} is backwards for one of the trends]

\ion{C}{4} -- Unlike \citet{Richards:2003}, who found only slight trends in \civ, we see a very clear trend with \ion{C}{4} where redder quasars have stronger (higher equivalent width) emission.  We attribute this difference between our work and that of \citet{Richards:2003} as being due to our better ability to separate dust from intrinsic redness (specifically dusty blue objects likely affected the reddest spectra in \citet{Richards:2003}) which is consistent with bluer objects having weaker \ion{C}{4}.  There is also a small blueward shift of the line for objects with weaker \ion{C}{4}, consistent with previous findings \citep[e.g.,][]{Sulentic:2007,Richards:2011,Gaskell:2013}, best seen here in the red wing of the line (as compared to the blue wing).
%As the spectra become bluer the strength of \ion{C}{4} decreases and the line becomes more blue shifted with respect to the quasars restframe. 

\ion{He}{2} -- The EW of the \ion{He}{2} $\lambda$1640\,\AA\ increases as the spectra become redder.  The same trend can be seen in the \ion{O}{3}]/\ion{Al}{2}/\ion{Fe}{2} blend just redward of \ion{He}{2}. \citet{Leighly:2004a} suggests that a stronger \ion{He}{2} emission line is an indicator of a harder (bluer) continuum in the X-ray. These trends are consistent with \citet{Richards:2003} \citep[see also][]{Baskin:2013}.

\ion{C}{3}] -- A clear trend of increasing EW with increasing redness is seen in the \ion{C}{3}] $\lambda$1909\,\AA\ line, consistent with \citet{Richards:2003}. The opposite trend is seen in the nearby \ion{Al}{3} $\lambda$1857\,\AA\ line (blue being stronger) and no trend is seen in the \ion{Si}{3}] $\lambda$1892\,\AA\ line, whereas \citet{Richards:2003} showed no trend with either line.  The combination of both strong \ion{Al}{3} and \ion{Si}{3}] with respect to \ion{C}{3}] in the bluer quasars is an indication that they have a softer (redder) ionizing spectrum as compared to the redder quasars \citep[See Appendix A3 of][]{Casebeer:2006,Grupe:2010}.

\ion{Mg}{2} -- There is a slight trend in \ion{Mg}{2} where the line get weaker and narrower as redness increases. As with H$\beta$, the width of \ion{Mg}{2} is proportional to the mass of the central black hole \citet[e.g.,][]{Peterson:2011}, possibly indicating that the bluer quasars have larger $\mbh$.

[\ion{O}{2}] -- This line shows the opposite trend as \ion{Mg}{2}, the line gets stronger (higher EW) and wider as redness increases, broadly consistent with \citet{Richards:2003}. % [Have coauthors that know more about this comment on possible SF contributions here.]
The combination of stronger [\ion{O}{2}] and weak Balmer line emission in the red quasars may indicate an underestimation in the host galaxy subtraction for these quasars \citep[e.g.,][]{Ho:2005}. For example, we have assumed a $z$=0 template for all quasars, which may not be appropriate.  However, for the difference to be due to host galaxy, the host galaxy properties would actually have to be different between the blue and red objects (and not just a function of redshift).
The stronger [\ion{O}{2}] may also indicate that the red quasars have more star formation in their host galaxies \citep{Rosa-Gonzalez:2002} in addition to the quasar's contribution.
 As such these differences do appear to reveal something about either the quasar or host galaxy physics.

H$\gamma$ and [\ion{O}{3}] -- The strengths of both H$\gamma$ and [\ion{O}{3}] $\lambda$4363\,\AA\ decrease as redness increases.

H$\beta$ -- The strength and width of the H$\beta$ decreases as redness increases. The width of this line has been shown to be an indicator of inclination, such that wider H$\beta$ corresponds to more edge-on views of the accretion disk \citep{Wills:1986,Shen:2014}.  Additionally the FWHM of H$\beta$ is used to estimate the virial velocity of the BLR when estimating black hole masses \citep[e.g.][]{Vestergaard:2006}, where a larger FWHM corresponds to a more massive black hole. However, if the host galaxy is underestimated preferentially in the red objects that could also bring the FWHMs into alignment.
The nearby [\ion{O}{3}] lines show a slight trend in the opposite direction, which could be an indication that the electron temperature of the emitting gas could be changing in response to the quasars continuum. 

H$\alpha$ and [\ion{N}{2}] -- As with the other Balmer lines, H$\alpha$ gets weaker as redness increases.

SBB -- The bottom panel of Figure~\ref{fig:ind_spec} shows a clear difference in the SBB region between the red and blue quasars with the SBB appearing weaker (lower EW) as redness increases, consistent with \citet{Richards:2003}. This is likely caused by a weakening of the Balmer emission as quasars become redder, consistent with the weakening of the resolved Balmer lines. 

Overall we find that the intrinsically redder quasars have larger EW high-ionization emission lines, but smaller EW low-ionization emission lines.  If the redder objects have harder continua in the EUV part of the spectrum, then all else being equal, we would expect both sets of lines to be stronger.  There are a number of possibilities for this difference.  First it could be that the BELR has multiple components: one dominated by high-ionization emission, the other by low-ionization emission \citep[e.g.,][]{Leighly:2004a,Collin:2006,Richards:2011} and those components have different covering fractions related to the shape of the SED.  Second could simply be that, if the lines have the same integrated flux, the redder objects with relatively lower continuum in the UV (where the high-ionization lines reside) will have larger EW high-ionization lines.  Lastly, uncorrected host galaxy light could be artificially inflating the continuum in the optical, making the low-ionization lines appear weaker in terms of EQW.

\subsubsection{Trends with E(B-V)}

Still examining the non-BAL spectra, we also examined the spectral lines while holding $\dal$ fixed and comparing spectra with different value of $\ebv$.  The clearest trend is seen in \ion{C}{4}, where the strength of the line decreases as the reddening increases.  Unlike \citet{Richards:2003} we do not see any increase in the strength of [\ion{O}{2}] with reddening, but we do see a slight increase in the [\ion{O}{3}] lines. The strong increase in [\ion{O}{2}] seen in \citet{Richards:2003} could have been caused by intrinsically red quasars being mistaken as dust reddened in their sample.   None of the other lines show any significant changes, thus we have not included a figure similar to Figure~\ref{fig:ind_spec} but as a function of $\ebv$.

%When studying the same spectral lines when holding the intrinsic color fixed and comparing different value for $\ebv$ find very different results than \citet{Richards:2003}.
%In Figure~\ref{fig:ind_spec2} we show the same spectral lines, this time holding the intrinsic color fixed ($0.05<\dal<0.38$) and showing different values for $\ebv$.  Like in Figure~\ref{fig:ind_spec} we choose a subset from each bin such that the $L_{2500\,\AA}$ distribution is the same for each spectra\footnote{We used the observed luminosity not corrected for reddening}. 
%The clearest trend is seen in \ion{C}{4}, where the strength and blueshift of the line both decrease as the reddening increase.  Unlike \citet{Richards:2003} we do not see any increase in the strength of [\ion{O}{2}] with reddening, but we do see an increase in the [\ion{O}{3}] lines.  The strong increase in [\ion{O}{2}] seen in \citet{Richards:2003} could have been caused by intrinsically red quasars being mistaken as dust reddened in their sample.

\subsubsection{Trends in BAL Spectra}

\begin{figure}[t]
\begin{center}
\includegraphics[width=3in]{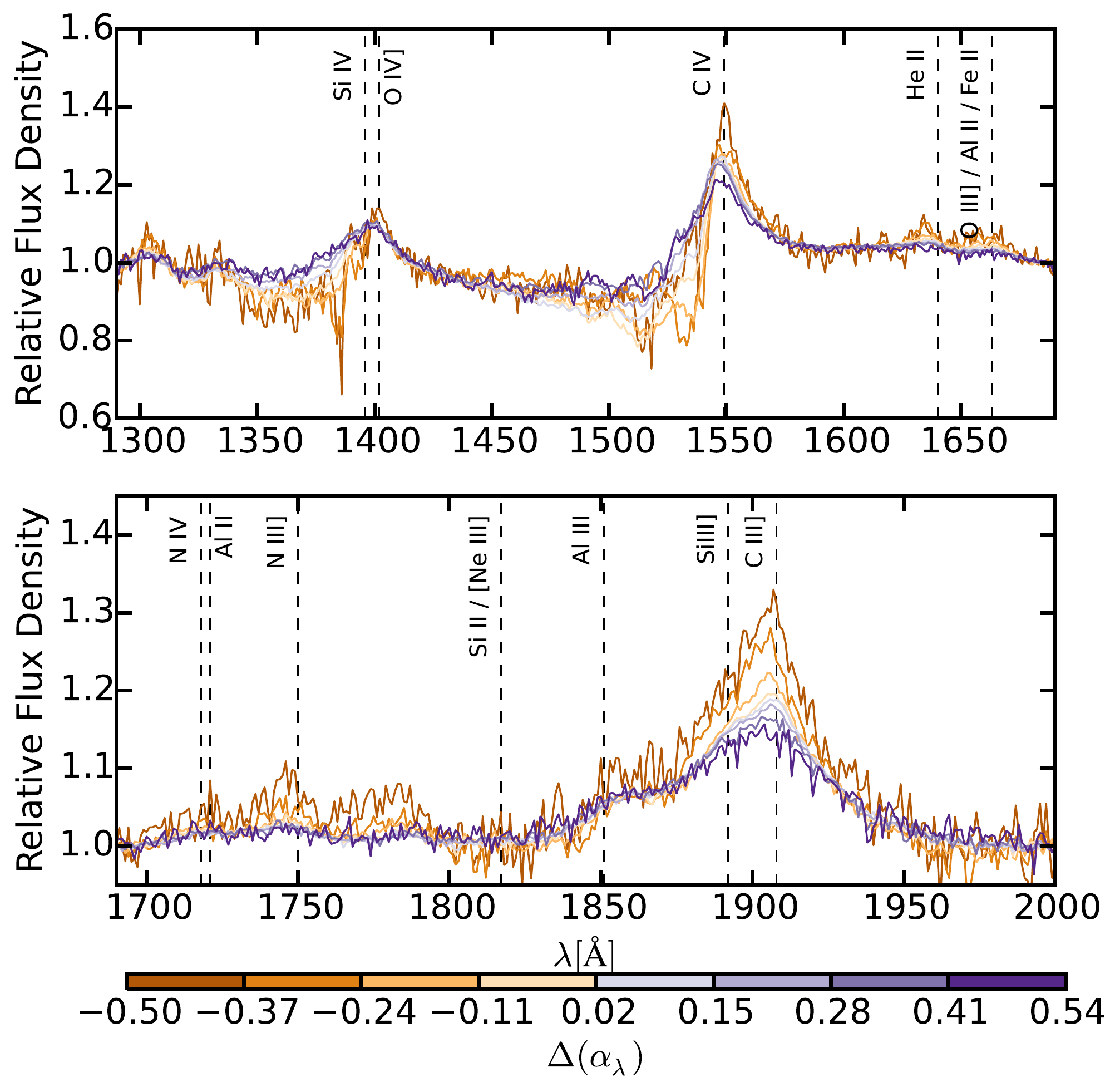}
\caption[Expanded emission-line regions for BAL quasars]{\label{fig:ind_bal_spec} Expanded emission-line regions from the lowest $\ebv$ bin from Figure~\ref{fig:bal_spec} ($0.012<\ebv<0.044$). The colors represent the $\dal$ value for each spectrum.  The spectra are normalized such that their continua agree at the edges of the panels. Intrinsically redder BALs have stronger \ion{C}{4} and \ion{C}{3}] emission lines.  The correlation see between changes in both the emission and absorption lines indicate they are both responding to changes in the underlying SED.}
\end{center}
\end{figure}

Looking instead at the BAL spectra we confirm differences in the emission and (mean) absorption as a function of (intrinsic) color (see Figure~\ref{fig:ind_bal_spec}), with the (intrinsically) redder objects having more strongly peaked \ion{C}{4}, stronger \ion{C}{3} (and \ion{Si}{3}) and some excess at $\sim$1750\,\AA, as was seen in Figure 9 of \citet{Reichard:2003b}. The excess near 1750\,\AA\ may be consistent with \ion{N}{3}] at 1750\,\AA\ and \ion{Fe}{2} UV 191 near 1788\,\AA\ \citep{Vanden-Berk:2001,Leighly:2007}. The absorption and emission are correlated as might be expected if both are responding to differences in the SED.  If the absorption was not responding to differences in the SED, the differences in the absorption troughs would be averaged away and little variation would be seen in the composites. Therefore, the correlation that we see indicates there may be some underlying trend relating both the absorption and emission \citep[e.g.,][]{Turnshek:1988, Baskin:2013}.  The narrow, low-velocity absorption seen in the redder BAL quasars (just blueward of the \ion{C}{4} and \ion{Si}{4} lines) could be an indication of weaker wind or a higher column density.  These quasars also show evidence of strong, not blueshifted \ion{C}{4} and strong \ion{C}{3}]---two properties that \citet{Richards:2011} argue are also indicators of a weaker winds.

%We note that radio-loud (RL) ``BALs'' \citep[e.g.,][]{Becker:2000} do not always meet the formal definition of BALs, having relatively weak absorption troughs at low velocity. As the RL quasars tend to be (on average) redder than radio-quiet (RQ) quasars, our finding of mini-BAL-like residuals in the reddest composites may not be that surprising. 

Our finding of residual mini-BAL troughs in the reddest composites is of interest in terms of the parent population of radio-loud BALs \citep[e.g.,][]{Becker:2000,Rochais:2014}.  Radio-loud quasars tend to be redder (on average) than radio-quiet quasars and their BALs troughs do not always meet the formal definition of BALs.  As such our findings may be consistent with the properties of RL BALs being driven by a somewhat different SED than traditional BALQSOs.

%[RL quasars also intrinsically redder.  So this could be consistent with the BALs in the redder objects being mini-BALs as opposed to troughs that meet the classical BAL definition.]
%[cmk: not sure I follow the connection between RL and mini-BALs]

\section{Conclusions} \label{sec:conclusions}

Using mid-IR thorough UV photometry alone we calculated intrinsic powerlaw slopes $\dal$ and SMC reddening law coefficients $\ebv$ for $\sim$35,000 uniformly selected quasars from the SDSS DR7 quasar catalog.  By using a hierarchical Bayesian model we were able to break the degeneracy between these parameters and keep both of them in a physical range (Figure~\ref{fig:indv_fits_SMC}). From these fits we found that:
\begin{enumerate}
\item BAL quasars are biased (blueward) within the parent population as compared to the non-BAL quasars.
%that non-BAL and BAL quasars do not have the same parent population.  
Both the BAL and non-BAL quasars are bluer than the modal spectral index ($\alpha_\lambda = -1.83$ and -1.79 respectively) while the BAL quasars have higher amounts of reddening (13\% with $\ebv>0.1$) than the non-BAL quasars (2.5\% with $\ebv>0.1$).
%The BAL quasars are slightly bluer ($\alpha_\lambda = -1.83$) and have higher amounts of reddening (13\% with $\ebv>0.1$) than the non-BAL quasars ($\alpha_\lambda = -1.79$, 2.5\% with $\ebv>0.1$).  
\item The BAL fraction is very dependent on the intrinsic color and amount of dust reddening of the quasars.  After marginalizing over $\ebv$ we find the {\em observed} BAL fraction increases from $\sim$0.05 to $\sim$0.2 as quasars become bluer (Figure~\ref{fig:bal_frac}). 
%The conjunction of high BAL fraction with extremely blue intrinsic continua and large $\ebv$ values implies that in these objects the BAL wind covers a large fraction of the sky.
%[Still something fishy there.   How many BALs and non-BALs are you studying? Also need to put back the objects with absorption lines when doing this.]
\end{enumerate}

We constructed composite spectra by splitting the resulting measurements of $\dal$ and $\ebv$ into a 10x10 grid. With the exception of the non-BAL quasars with the most reddening and the bluest quasars with the least reddening, the fit values for each bin do a good job of reproducing the slopes and curvatures of the spectra, showing that our fit values are consistent with being physically meaningful (Figures~\ref{fig:spec} and~\ref{fig:bal_spec}).  Comparing the spectra in the lowest reddening bin revealed trends in the spectral lines based on a quasar's intrinsic color.
Exploring the composite spectral revealed that:
\begin{enumerate}
\item Within the BAL sample the \ion{C}{4} absorption trough becomes deeper as $\ebv$ increases and becomes wider as quasars become bluer.
%Exploring the spectra further, we found that the highly reddened BAL quasars are dominated by LoBALs while the less reddened sample is dominated by HiBALs.  This trend is in agreement with many previous studies \citep[e.g., ][]{Weymann:1991,Reichard:2003a,Reichard:2003b}, and is consistent with the orientation-dependent picture presented in \citet{Weymann:1991}.
\item Intrinsically red quasars have stronger \ion{C}{4}, \ion{He}{2}, \ion{C}{3}], and [\ion{O}{2}], while the \ion{Mg}{2}, H$\gamma$, H$\beta$, H$\alpha$, and the SBB are all weaker (Figure~\ref{fig:ind_spec}).  
%The stronger UV emission lines in the intrinsically red quasars indicate that they have more ionizing photons (harder SED) in the energy range $\sim$13.6--64.5\,eV (15.5$\lesssim\log{(\nu [Hz])}\lesssim$16.2) than the (optically) bluer quasars.  
In Section~\ref{sec:indv_lines} we discussed three scenarios that could explain the difference between the strengths of the high-ionization emission lines and the low-ionization emission lines: a two-component BELR, comparison of EQW rather than line luminosity, and host galaxy contamination.  It is further possible that the larger width of the H$\beta$  Hb and \ion{Mg}{2} lines in the bluer quasars could indicate that they have larger $\mbh$ or that they might instead be viewed closer to edge on.
%Additionally, the width of H$\beta$ and \ion{Mg}{2} are proportional to the mass of the central black hole, indicating that the bluer quasars likely have larger $\mbh$, or that they 
%.  Or if H$\beta$ is used as an indicator of inclination, the bluer quasars 
%might be instead viewed closer to edge-on.
%or have host galaxy contamination that is a function of color.  
%We will take a closer look at these trends in Chapter~\ref{BH}.
%When holding the intrinsic color fixed and looking at spectra with different amounts of reddening, we only see trends in \ion{C}{4} (weaker for more reddened quasars) and [\ion{O}{3}] (stronger for more reddened quasars). 
\item In the BAL sample similar trends were found in \ion{C}{3}] and \ion{C}{4} (stronger for red-continuum quasars), and we additionally found that the absorption troughs are similarly responding to the changes in the underlying SED (Figure~\ref{fig:ind_bal_spec}).
%that the depth of the \ion{C}{4} and \ion{Si}{4} BAL troughs are more consistent for redder quasars (Figure~\ref{fig:ind_bal_spec}).
\end{enumerate}

\acknowledgments

G.T.R. acknowledges support from an Alfred P. Sloan and an Alexander von Humboldt Research Fellowship along with NASA grants NNX08AJ27G, NNX10AF74G, and NNX12AI49G 
and the generous support of a research fellowship from the Alexander von Humboldt Foundation at the Max-Planck-Institut f{\"u}r Astronomie and is grateful for the hospitality of the Astronomisches Rechen-Institut.  S.C.G. thanks the Natural Science and Engineering Research Council of Canada for support.
We thank both David Schiminovich and David Hogg for providing {\em GALEX} forced photometry along with Anna Sajina, Jonathan Stern, Erica Smith, and Hagai Netzer for their comments on the manuscript.
This publication makes use of data products from the {\em Wide-field Infrared Survey Explorer}, which is a joint project of the University of California, Los Angeles, and the Jet Propulsion Laboratory/California Institute of Technology, funded by the National Aeronautics and Space Administration. 
Funding for the SDSS and SDSS-II has been provided by the Alfred P. Sloan Foundation, the Participating Institutions, the National Science Foundation, the U.S. Department of Energy, the National Aeronautics and Space Administration, the Japanese Monbukagakusho, the Max Planck Society, and the Higher Education Funding Council for England. The SDSS Web Site is \url{http://www.sdss.org/}.
All figures made use of matplotlib\footnote{\url{http://matplotlib.org}} \citep{matplotlib} and Figures 1 and 5-11 were made using densityplot\footnote{\url{https://github.com/CKrawczyk/densityplot}} \citep{Krawczyk:2014}.

\bibliography{All_refs}
\appendix
\section{Bayesian Model} \label{ch:mcmc}

In a Bayesian framework, the fit parameters for a model are not assumed to be fixed in value, but instead, these parameters are assumed to come from some probability distribution.  Bayes' theorem tells us how to express this probability distribution in terms of the likelihood of the {\em data} given a set of model parameters and any prior information we have about the model parameters.  Applying Bayes' theorem to a toy model that has a set of $n$ model parameters, $\vec{\phi} = \{\phi_0,\phi_1,...,\phi_n\}$, and measured data, $D$, gives:
\begin{equation}
 P(\vec{\phi}|D) \propto P(\vec{\phi})P(D|\vec{\phi})
\end{equation}
or in words: the posterior distribution for $\vec{\phi}$ is proportional to the prior distribution of $\vec{\phi}$ times the likelihood of the data, given $\vec{\phi}$.

In hierarchical Bayes, the prior distribution, $P(\vec{\phi})$, is not known or assumed beforehand, but instead is a function of another set of $m$ parameters, $\vec{\theta} = \{\theta_0,\theta_1,...,\theta_m\}$, known as hyperparameters.  This leads to the joint posterior distribution for $\vec{\phi}$ and $\vec{\theta}$ of:
\begin{equation}
 P(\vec{\theta},\vec{\phi}|D) \propto P(\vec{\theta})P(\vec{\phi}|\vec{\theta})P(D|\vec{\phi})
\end{equation}
where $P(\vec{\theta})$ are the hyperpriors that are placed on $\vec{\theta}$ \citep{Gelman:2006}.

With the posterior distribution defined by Bayes' theorem, MCMC sampling methods can be used to simulate data points drawn from it, providing the full joint probability distribution for all parameters and hyperparameters given the observed data. When using a hierarchical model, care must be taken to make sure that the sampling method is self consistent; that is, when $\vec{\phi}$ changes, $\vec{\theta}$ changes to match the new distribution of $\vec{\phi}$.  This can be achieved by using a Gibbs sampler \citep{Gelman:2006}.

\subsection{Gibbs Sampling}
In Gibbs sampling, sample points are drawn from the posterior by drawing from the {\em conditional probabilities} for each parameter (or set of parameters) in turn. These conditional probabilities are found by taking the posterior distribution and holding all but one (or a small set of) parameters fixed.  As an example, take a model that has two parameters, $A$ and $B$, being estimated. Given inital values, $A_0$ and $B_0$, the Gibbs sampler updates these values by alternating between drawing a new $A$ value from the posterior holding $B$ fixed, $A_{i+1} \sim P(A|B_i,D)$, and then drawing a new $B$ value from the posterior holding $A$ fixed at its new value, $B_{i+1} \sim P(B|A_{i+1},D)$, where ``$\sim$'' is read ``distributed as.'' After a sufficient number of iterations, known as the burn-in period, each step of the Gibbs sampler will simulate a draw from the joint distribution $P(A,B|D)$.

For some models these conditional probabilities have known distributions. One such model is estimating the mean, $\mu$, and standard deviation, $\sigma$, when given a set of $n$ normally distributed observations $\vec{x}$. The conditional probabilities from this model are:
\begin{eqnarray}
 P(\mu|\sigma,\vec{x}) \sim \operatorname{N}\left( \frac{\sum_i^n x_i}{n}, \frac{\sigma}{\sqrt{n}} \right) \\
 P(\sigma^2|\mu,\vec{x}) \sim \frac{(n-1)\sum_i^n(x_i-\mu)^2}{n}\frac{1}{\chi_{n-1}^2} 
\end{eqnarray}
where $\operatorname{N}(a,b)$ is a normal distribution with mean $a$ and standard deviation $b$, and $\chi_{n-1}^2$ is a chi-square random variable with $n-1$ degrees of freedom \citep{Gelman:2006}.

When the conditional probability cannot be represented by a known distribution, draws can still be simulated by using another MCMC sampler. Typically, a Metropolis-Hasting \citep[MH;][]{Hastings:1970} step is used, resulting in a method called Metropolis-within-Gibbs sampling \citep{Metropolis:1953}, but {\em any} MCMC sampler can be used to draw from the conditional.  In this work, we use the affine invariant MCMC ensemble sampler \texttt{emcee}, described in \citet{Foreman-Mackey:2013}\footnote{Python code available at \url{http://dan.iel.fm/emcee/current/}}, to form an \texttt{emcee}-within-Gibbs sampler.  This sampler efficiently samples distributions with degenerate variables, and it only has one tuning parameter, unlike the MH sampler, which has one tuning parameter for each fit parameter and is slow to converge for degenerate variables.

The \texttt{emcee} sampler works by simulating $N$ simultaneous draws from the conditional distribution, where each draw is known as a walker. The proposal distribution for each walker is based on the current positions of all the remaining walkers. In each step, a new position is proposed for each walker by``walking" it along the vector connecting it to another randomly selected walker. If the likelihood of the conditional is higher at this proposed value, the step is accepted; if it is not, it is accepted with a probability related to the ratio of the likelihoods at its current position and the proposed position (see \citet{Foreman-Mackey:2013} for more details). Because this sampler uses an ensemble of walkers, at every step of the algorithm it produces $N$ estimates for each parameter.

To form an \texttt{emcee}-within-Gibbs sampler some care must be taken in choosing the number of walkers, $N$, needed.  The general rule of thumb is $N$ should be larger then twice the number of parameters being estimated.  With hierarchical models, the number of parameters can grow to be very large.  If every parameter was estimated in the same step (i.e. when not using Gibbs sampling) then $N$ would be too large to make the \texttt{emcee} sampler a practical solution.  When placing the \texttt{emcee} sampler within a Gibbs step, the number of parameters being estimated in each conditional is small (3 for the model considered here), meaning $N$ only needs to be larger than twice the dimensionality of the largest conditional, bringing $N$ down to a reasonable size.

%In a typical Gibbs step, only a small set of parameters are stepped forward while the others are held fixed at their current value. As mentioned above, \texttt{emcee} produces $N$ estimates for each parameter in every step, unlike a Gibbs or MH sampler, where only one estimate is produced. Given the way the \texttt{emcee} algorithm steps forward each walker, the same value for each of the fixed parameters must be used when evaluating the likelihood of each walker. This raises the question: what value should be chosen to represent the fixed set of parameters in each conditional? For our sampler, we pick one of the walkers at random to represent the fixed parameters in each conditional. [I have no idea if this the propper way to deal with this.  I have also tried using the median vlaue of all the walkers as the fixed values, but that seems to make the convergence rate worse...   GTR: Ask Hogg, Daniel?] When updating a parameter that can be represented by a known distribution (i.e. \texttt{emcee} is not used), we treat each walker as an {\em independent} Gibbs chain.  In other words, the fixed parameters are taken to be each walker's current position.

\subsection{The Model} \label{sec:mcmc_model}

For this paper we create a hierarchical model we use to estimate the slopes and curvatures for each quasar in our sample.
We start by modeling each observed $\mrel_{ij}$ as coming from a Student's $t$-distribution with 3 degrees of freedom %normal distribution 
centered on the model given by Equation~\ref{eqn:rel_red_eq}(b), with components $\vec{m_0}$, $\vec{\dal}$, and $\vec{\ebv}$, and with a width given by the observed measurement error, $\sigma_{ij}$, combined with an unknown term $S$ representing the error in the modal magnitudes:
\begin{equation} \label{eqn:model}
 \mrel_{ij} \sim \operatorname{T}\left(3,\mrel_{ij}^{\rm model},\sqrt{\sigma_{ij}^2+S^2}\right),
\end{equation}
where the $t$-distribution is:
\begin{equation}
 \operatorname{T}(\nu,\mu,\sigma) \propto \frac{1}{\sqrt{\nu \pi}\sigma} \left( 1 + \frac{\mu^2}{\sigma^2 \nu} \right)^{-(\nu+1)/2},
\end{equation}
$\mrel_{ij}^{\rm model}$ is the model for $\mrel$ given in Equation~\ref{eqn:rel_red_eq}(b), $S$ can be thought of as a ``calibration'' error in our model, $i$ is used to index each quasar, and $j$ is used to index each filter. We use a Student's $t$-distribution instead of a normal distribution to make our model more robust against outliers. A value of $\nu=3$ degrees of freedom implies $\sim6\%$ of the data are outliers by more than 3$\sigma$ \citep{Kelly:2012}.

With the likelihood of our data given the model defined, we next model the prior distributions for our set of fit parameters $\phi=\{ \vec{m_0}, \vec{\dal}, \vec{\ebv} \}$.  The collection of powerlaw slopes, $\vec{\dal}$, are modeled as coming from a normal distribution with unknown mean, $\mu_{\alpha}$, and standard deviation, $\sigma_{\alpha}$. This distribution was chosen since we expect to see just as many intrinsically blue as intrinsically red quasars.  Unlike \citet{Hopkins:2004} we do not assume $\mu_{\alpha}$ is zero; this way we account for differences between the {\em observed} modal quasar and the necessarily bluer {\em intrinsic} modal quasar (i.e. the modal quasar after correcting for dust reddening). We additionally limit the values of $\vec{\dal}$ to stay in the range $[-0.5,2.3]$, where the lower limit is equivalent to a quasar with $\alpha_{\lambda} \sim -1.22$ (a very red quasar) and the upper limit is equivalent to a quasar with $\alpha_{\lambda} \sim -4$ (an infinite temperature blackbody).

The collection of reddening amounts, $\vec{\ebv}$, are modeled as coming from an EMG with unknown shape parameters. This distribution is the result of summing together random variables drawn from a normal distribution with mean $\mu_{\rm dust}$ and standard deviation $\sigma_{\rm dust}$, and random variables drawn from a exponential distribution with rate parameter $\lambda_{\rm dust}$\,\footnote{Smaller values of $\lambda_{\rm dust}$ indicate a heavier tail.} (i.e., the convolution of a normal distribution with and exponential distribution). This is similar to the half-normal, half-exponential distribution \citet{Hopkins:2004} found to work well for quasars. As with the $\vec{\dal}$ distribution, we do not assume $\mu_{\rm dust}$ is zero beforehand.

To normalize our data, we have set the $i$ band $\mrel$ to zero for all quasars. Under this normalization the offset parameters, $\vec{m_0}$, will be normally distributed with zero mean and a standard deviation of $\sqrt{\sigma_{i{\rm  band}}^2+S^2}$. For our data the observed $i$ band errors are small compared to $S$, allowing us to drop the $\sigma_{i{\rm  band}}$ term.

Using these distributions, our priors are:
\begin{eqnarray} \label{eqn:priors}
 m_{0,i} &\sim& \operatorname{N}(0,S) \nonumber \\
 \dal_{i} &\sim& \operatorname{N}(\mu_{\alpha},\sigma_{\alpha}) \\
 \ebv_{i} &\sim& \operatorname{N}(\mu_{\rm dust},\sigma_{\rm dust}) + \operatorname{E}(\lambda_{\rm dust}) \nonumber 
\end{eqnarray}
where $\operatorname{E}(\lambda)$ is an exponential distribution with rate parameter $\lambda$.
To complete the definition of our model, we place uninformative hyperpriors on the hyperparameters $\theta=\{ \mu_{\alpha},\sigma_{\alpha},\mu_{\rm dust},\sigma_{\rm dust},\lambda_{\rm dust},S \}$:
\begin{eqnarray} \label{eqn:h_priors}
 \mu_{\alpha} &\sim& \operatorname{U}(-2,2) \nonumber \\
 \sigma_{\alpha} &\sim& \operatorname{U}(0,4) \nonumber \\
 \mu_{\rm dust} &\sim& \operatorname{U}(-5,5) \\
 \sigma_{\rm dust} &\sim& \operatorname{U}(0,10) \nonumber \\
 \lambda_{\rm dust} &\sim& \operatorname{U}(0,40) \nonumber \\
 S &\sim& \operatorname{U}(0,10) \nonumber
\end{eqnarray}
where $\operatorname{U}(a,b)$ is a uniform distribution between the values $a$ and $b$.

As written, equation~\ref{eqn:model} is dependent on both $\phi$ and $\theta$.  Leaving it like this can lead to instabilities in our sampler.  These instabilities can be removed to introducing a set of nuisance parameters, $\delta_{ij}$, into $\phi$ that represent the normally distributed noise introduced by $S$. Equation~\ref{eqn:model} now becomes:
\begin{equation} \label{eqn:model2}
 \mrel_{ij} \sim \operatorname{T}(3,\mrel_{ij}^{\rm model}+\delta_{ij},\sigma_{ij})
\end{equation}
and one more prior is added to equation~\ref{eqn:priors}:
\begin{equation}
 \delta_{ij} \sim \operatorname{N}(0,S)
\end{equation}

\subsection{Conditional Likelihoods} \label{sec:conditionals}
To make use of a Gibbs sampler to estimate both $\phi$ and $\theta$, the conditional likelihoods for each parameter (or set of parameters) need to be found. Taking the model defined in Appendix~\ref{sec:mcmc_model}, several of the fit parameters have conditionals with analytic forms:
\begin{eqnarray}
 \mu_{\alpha} &\sim& \operatorname{N}\left( \frac{\sum_i^n \dal_i}{n},\frac{\sigma_{\alpha}}{\sqrt{n}} \right) \\
 \sigma_{\alpha}^2 &\sim& \frac{(n-1) \sum_i^n (\dal_i-\mu_{\alpha})^2}{n} \frac{1}{\chi_{n-1}^2} \\
 S^2 &\sim& \frac{\sum_i^n \sum_j^m (\delta_{ij}+m_{0,i})^2}{2} \frac{1}{\chi_{nm}^2} \\
 \delta_{ij} &\sim& \operatorname{N}\left( \frac{S^2}{S^2+\sigma_{ij}^2} (\mrel_{ij}-\mrel_{ij}^{\rm model}), \frac{S^2 \sigma_{ij}^2}{S^2+\sigma_{ij}^2} \right)
\end{eqnarray}

For the remaining conditionals it is useful to define the probability density functions for the EMG distribution and the normal distribution:
\begin{eqnarray}
 \operatorname{EMG}(x|\mu,\sigma,\lambda) &=& \frac{\lambda}{2} \exp{\left(\frac{\lambda}{2} (2\mu+\lambda \sigma^2-2x) \right)} \nonumber \\
 & &\operatorname{erfc}\left( \frac{\mu+\lambda \sigma^2-x}{\sqrt{2} \sigma} \right) \\
 \operatorname{N}(x|\mu,\sigma) &=& \frac{1}{\sqrt{2\pi}\sigma} \exp{\left(\frac{-(x-\mu)^2}{2\sigma^2} \right)} 
\end{eqnarray}
where $\operatorname{erfc}$ is the complementary error function. The remaining conditional probabilities are:
\begin{eqnarray}
 &P&(\mu_{\rm dust},\sigma_{\rm dust},\lambda_{\rm dust}|\ebv) \propto P(\mu_{\rm dust})P(\sigma_{\rm dust}) \nonumber \\
& &P(\lambda_{\rm dust}) \prod_i \operatorname{EMG}(\ebv_i|\mu_{\rm dust},\sigma_{\rm dust},\lambda_{\rm dust})  \label{eqn:mcmc_data_eqn11} \\
 &P&(\phi_i|\theta,\vec{\mrel_i}) \propto \operatorname{EMG}(\ebv_i|\mu_{\rm dust},\sigma_{\rm dust},\lambda_{\rm dust}) \nonumber \\
& & \operatorname{N}(\dal_i|\mu_{\alpha},\sigma_{\alpha}) \operatorname{N}(m_{0,i}|0,S) \nonumber \\
& & \prod_j \operatorname{T}(\mrel_{ij}|3,\mrel_{ij}^{\rm model}+\delta_{ij},\sigma_{ij})  \label{eqn:mcmc_data_eqn12}
\end{eqnarray}
By breaking up the conditionals in this manner, the largest number of parameters being estimated in any given \texttt{emcee} step is 3.  As discussed in Section~\ref{sec:mcmc}, this means a few hundred walkers are sufficient to fully sample the space.

In order to increase the efficiency of our MCMC sampler, we use an Ancillarity-Sufficiency Interweaving Strategy (ASIS) similar to the one used in \citet{Kelly:2011} \citep[see also][]{Yu:2011}.  The ASIS works by introducing the following change of variables for Equation~\ref{eqn:mcmc_data_eqn12}:
\begin{eqnarray}
\tilde{\delta_{ij}} &=& \mrel_{ij}^{\rm model}+\delta_{ij}  \label{eqn:delta_trans} \\
\tilde{\delta_{ij}} &\sim& \operatorname{T}(3,\mrel_{ij}^{\rm model},S) 
\end{eqnarray}
This change allows for a second way of updating the individual fit parameters $\vec{m_{0}}$, $\vec{\dal}$, and $\vec{\ebv}$ that is not directly dependent on the observed data:
\begin{eqnarray} \label{eqn:mcmc_data_eqn2}
&P&(\phi_i|\theta,\tilde{\delta_{ij}}) \propto \operatorname{EMG}(\ebv_i|\mu_{\rm dust},\sigma_{\rm dust},\lambda_{\rm dust}) \nonumber \\
& & \operatorname{N}(\dal_i|\mu_{\alpha},\sigma_{\alpha}) \operatorname{N}(m_{0,i}|0,S) \nonumber \\
& & \prod_j \operatorname{T}(\tilde{\delta_{ij}}|3,\mrel_{ij}^{\rm model},S)
\end{eqnarray}
ASIS replaces the update step given by Equation~\ref{eqn:mcmc_data_eqn12} with the following steps:
\begin{enumerate}
    \item{Update $\vec{m_{0}}$, $\vec{\dal}$, and $\vec{\ebv}$ using Equation~\ref{eqn:mcmc_data_eqn12}}
    \item{Find $\tilde{\delta_{ij}}$ using Equation~\ref{eqn:delta_trans} and the current values for the fit parameters}
    \item{Update $\vec{m_{0}}$, $\vec{\dal}$, and $\vec{\ebv}$ using Equation~\ref{eqn:mcmc_data_eqn2}}
    \item{Find the new values for $\delta_{ij}$ associated with these parameters by apply the inverse of the transformation given in Equation~\ref{eqn:delta_trans}}
\end{enumerate}

In our analysis we used 200 walkers with a 2000 step burn-in and 400 steps of sampling.  In Section~\ref{sec:mcmc_results} we refer to $\theta$ as the population parameters and $\phi$ as the individual parameters.

\end{document}